\documentclass[]{raa}            

\usepackage{graphicx,times}             
\usepackage{natbib}
\usepackage{amssymb,amsmath}
\bibpunct{(}{)}{;}{a}{}{,}

\usepackage[a4paper=true,pagebackref=true]{hyperref}
\hypersetup{colorlinks = true, linkcolor = green, anchorcolor = red, citecolor = blue, filecolor = red, pagecolor = red, urlcolor = red}

\newcommand{\rad}{\ensuremath{\,{\rm rad}}}

\newcommand{\ns}{\ensuremath{\,{\rm ns}}}

\begin{document}

    \title{Reflections and Standing Waves on the Tianlai Cylinder Array}

    \volnopage{Vol.0 (20xx) No.0, 000--000}      
    \setcounter{page}{1}          

    \author{Jixia Li
        \inst{1,2}
    \and Fengquan Wu 
        \inst{1}
    \and Shijie Sun
        \inst{1,2}
    \and Zijie Yu
        \inst{1,2}
    \and Shifan Zuo 
        \inst{3,1,2}
    \and Yingfeng Liu
        \inst{1,2}  
    \and Yougang Wang
        \inst{1}    
    \and Cong Zhang
        \inst{1,2}
    \and Reza Ansari
        \inst{4}
    \and Peter Timbie
        \inst{5}
    \and Xuelei Chen
        \inst{1,2,6}
    }

   \institute{National Astronomical Observatories, Chinese Academy of Sciences,
             Beijing 100101, 
 China; {\it xuelei@cosmology.bao.ac.cn}\\
        \and
             School of Astronomy and Space Science, University of Chinese Academy of Sciences, Beijing 100049, China\\
        \and
             Center for Astrophysics and Department of Astronomy, Tsinghua University, Beijing 100084, China\\
        \and 
             IJC Lab, CNRS/IN2P3 \& Universit\'e Paris-Saclay, 15 rue Georges Cl\'emenceau, 91405 Orsay, France\\
        \and 
            Department of Physics, University of Wisconsin Madison, 1150 University Ave, Madison WI 53703, USA\\
        \and 
             Center of High Energy Physics, Peking University, Beijing 100871, China \\
\vs\no
   {\small Received~~20xx month day; accepted~~20xx~~month day}}

\abstract{In 21~cm intensity mapping, the spectral smoothness of the foreground is exploited to separate it from the much weaker 21~cm signal. However, the non-smooth frequency response of the instrument complicates this process.  Reflections and standing waves generate modulations on the frequency response. Here we report the analysis of the standing waves in the bandpass of the signal channels of the Tianlai Cylinder Array. By Fourier transforming the bandpass into the delay time domain, we find various standing waves generated on the telescope. A standing wave with time delay at $\sim$142~ns is most clearly identified which is produced in the 15-meter feed cable. We also find a strong peak at a shorter delay of $\tau < 50 \ns$, which may be a mix of the standing wave between the reflector and feed, and the standing wave on the 4~m intermediate frequency (IF) cable. We also show that a smoother frequency response could be partially recovered by removing the reflection-inducted modulations. However, the standing wave on the antenna is direction-dependent, which poses a more difficult challenge for high precision calibration. 
}

   \authorrunning{Jixia Li et al.}            
   \titlerunning{Analysis of Tianlai Cylinder Array}  

   \maketitle

%
%

\newcommand{\red}[1]{\textcolor{red}{{#1}}}
\newcommand{\blue}[1]{\textcolor{blue}{{#1}}}

\section{Introduction}           
\label{sec:introduction}

The Tianlai (literally ``heavenly sound'' in Chinese) project is a 21~cm intensity mapping experiment \citep{2012IJMPS..12..256C,chen2015tianlai,chen2015AAPPS,PAON4_Zhang_2016,2016RAA....16..158Z}. The pathfinder experiment includes a cylinder array and a dish array, both located at the Hongliuxia site ($91^\circ 48^\prime \mathrm{E}, 44^\circ 09^\prime \mathrm{N}$), in Xinjiang, northwest China \citep{wu2014site}. The aim of the experiment is to test the principle and key technologies for conducting large scale structure surveys of the neutral hydrogen distribution in the redshift range of $0\sim 3$. If successful, the experiment can be expanded to larger scale, and the full scale experiment could provide measurement of the dark energy equation of state parameters and constrain inflationary features \citep{Xu2015,Xu2016}. The two arrays had their first light observation in 2016 \citep{Das2018,Li2020,Wu2020}.

21~cm intensity mapping is an efficient way to conduct low angular resolution tomographic surveys of neutral hydrogen (HI), such that the individual galaxies are not resolved, but the large scale structure is mapped \citep{Kovetz2019}. The technique 
is considered to be a potentially very powerful cosmological probe: it could be applied to measuring the equation of state of dark energy from the baryon acoustic oscillation (BAO) features (e.g. \citep{Xu2015}), detecting inflationary features  (e.g. \citep{Xu2016}), as well as studying the Epoch of Reionization (EoR). A variety of experiments, ranging from those for the EoR (e.g. LOFAR \citep{LOFAR2013}, MWA \citep{MWA2013}, PAPER \citep{PAPER2010}, HERA \citep{HERA2017}), to the post-EoR (such as the CHIME \citep{CHIME}, HIRAX \citep{HIRAX}, and BINGO \citep{BINGO}), are devoted to 21~cm tomographic surveys. Experiments have also been carried out with existing telescopes, such as the Green Bank Telescope (GBT) \citep{Masui2013,Switzer2013} and the Parkes Observatory \citep{Anderson2018}, and also proposed for large general purpose telescopes such as FAST \citep{Hu_2020} and SKA.

However, the 21~cm signal is $4\sim 5$ orders of magnitude smaller than the foreground radiation,  which includes galactic synchrotron, galactic free-free and extra-galactic radio sources (c.f. \citealt{Huang:2018ral}). Extracting the 21~cm signal generally relies on the fact that foreground emissions are smooth functions of frequency, while the 21~cm spectrum has a structure arising from the large-scale distribution of matter along the line of sight \citep{Liu&Shaw2019}. However, instrumental effects can introduce structures into the spectrum of otherwise smooth foregrounds. It is therefore very important to identify such instrument-induced modulations and remove them in order to detect the 21~cm signal. 

In this paper, we study the spectral modulations found in the auto-correlation visibilities of the Tianlai cylinder pathfinder. The components of the telescopes are all designed to have flat or smooth frequency responses. Nevertheless, we found that there are significant variations in their frequency responses. As we will show below, the responses have significant modulations which are probably caused by reflections and standing waves at the interfaces of different components of the systems. 

Although reflections and standing waves can be seen in every radio telescope, it has been discussed only in a few papers, e.g. \citet{Popping2007TheSW} for WSRT and \citet{Kern:2019ytc,Kern2019MitigatingII} for HERA. In the present paper, we analyze the standing waves on the Tianlai cylinder and discuss the possibility of its mitigation. 

\section{The Instrument}
\label{sec:instrument}

The Tianlai cylinder pathfinder array is described in detail in \citet{Li2020}. Here, we shall give a brief summary of the instrument. 

\begin{figure}
\centering
 \includegraphics[width=0.9\textwidth]{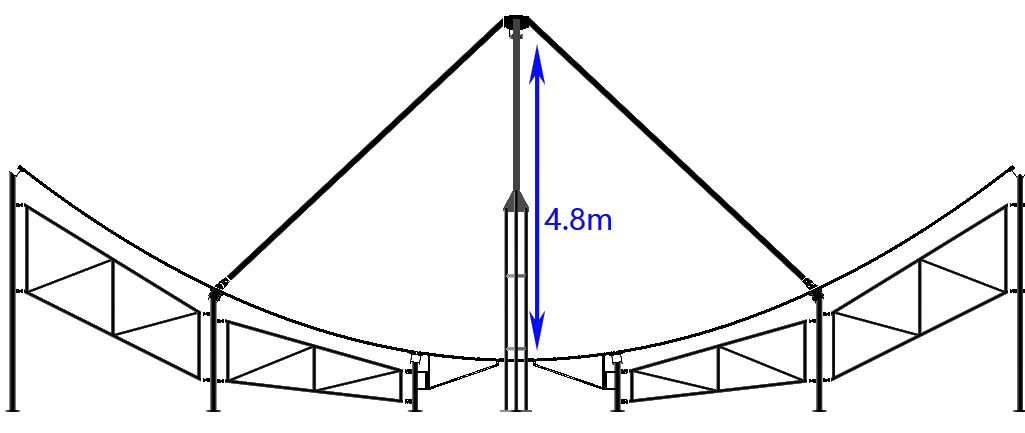}
  \caption{The schematic diagram of one reflector as viewed from south/north direction. The cylinder aperture is 15 m wide. The focal ratio is 0.32, so the distance between the feed and the reflecting parabolic surface is 4.8 m.}
 \label{fig:cylinder}
\end{figure}

The Tianlai cylinder array has three adjacent parabolic cylinder reflectors, each $40 {\rm m} \times 15 {\rm m}$ with its long axis oriented in the north-south (N-S) direction. The cylinders are closely spaced in the East-West (E-W) direction. The focal ratio of the cylinder is $f/D=0.32$, so the height of the feed is 4.8 meters above the reflector surface, as shown in Fig.~\ref{fig:cylinder}. Dual-polarization dipole feeds are placed along the focal line of each cylinder \citep{chen2016design}. The cylinder reflectors are fixed on the ground. At any moment, the instantaneous field of view (FoV) is a narrow strip running from north to south through the zenith. The latitude of the telescope site is $44^\circ$, which gives the declination of the points passing through the zenith. As the Earth rotates, it scans the northern celestial hemisphere. 

The three cylinders are denoted as cylinder A, B, C from east to west. Each has been installed with a slightly different number of feeds, 31, 32 and 33, respectively. From north to south, the feeds in each cylinder are labelled in numbers $1, 2, 3, ...$. The northernmost (southernmost) feeds A1, B1, C1 (A31, B32, C33) are aligned and the distance between the northernmost and southernmost feeds are 12.4 m. The unequal feed spacings on the three cylinders are designed to reduce the grating lobes \citep{2016RAA....16..158Z}. Each dual linear polarization feed generates two signal outputs. We will denote the N-S polarization as X and the E-W direction as Y. For example, the E-W polarized output of the 3rd feed in the middle cylinder will be referred to as B3Y.

\begin{figure}
\centering
\includegraphics[width=0.9\textwidth]{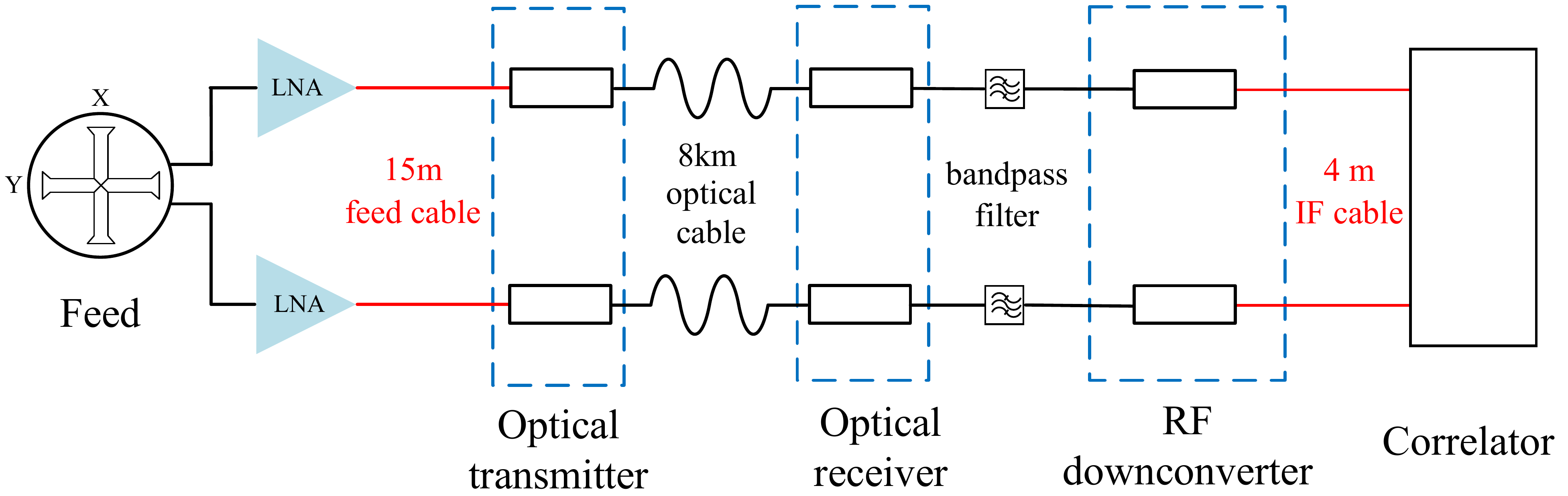}
 \caption{The schematic of the Tianlai analog signal chain. }
 \label{fig:schematic}
\end{figure}

The electric voltages induced by radio waves in the feed circuit are amplified by low noise amplifiers (LNAs) on the feed and sent via a 15 meter long coaxial cable (heretofore referred to as the {\it feed cable}) to the optical transmitter housed in the boxes below the antenna. The radio frequency (RF) electrical signal is then converted to an amplitude-modulated optical signal, which is then transmitted through an optical fiber to the station house, located about 6~km away in the nearby village of Hongliuxia. The optical signal is converted back to the RF electric signal in the analog electronics room, then down-converted to the intermediate frequency (IF).  Each of the IF signals is then transmitted through a 2 meter long coaxial cable to an SMA connector installed on a bulkhead mounted between the analog and digital electronics room. A similar 2 meter long coaxial cable on the digital room side passes the signal to the digital correlator, where it is digitized and processed. These cables will be referred to as the {\it IF cable}. A schematic of the analog part of the signal chain is shown in Fig.~\ref{fig:schematic}.

\begin{table}[htbp]
  \caption{Main design parameters of the Tianlai cylinder array.}
  \centering
  \begin{tabular}{l l}
    \hline
    Cylinders & $3 \times$ 15.0 m (EW) $\times$
    40.0 m (NS).\\
    Number of feeds & 96 (A:31, B:32, C:33) \\
    Feed spacing (cm)  & 41.33, 40.00, 38.75 \\
    f/D & 0.32 \\
    Feed illumination angle  &  $152^\circ$ \\
    Current frequency range and resolution & 700--800 MHz, 122 kHz \\
    X-pol(N-S) FWHM @750 MHz & $1.6^\circ$ (H-plane), ~$62.2^\circ$(E-plane)  \\
    Y-pol(E-W) FWHM @750 MHz & $1.8^\circ$ (E-plane), ~$71.4^\circ$(H-plane)  \\
    Location & E $91^\circ48'$, N $44^\circ9'$\\
    \hline
  \end{tabular}
  \label{tab:cylinder_parameter}
\end{table}

The whole system is designed to operate over a wide range of frequencies (400--1500~MHz), while the working frequency band is set by replaceable bandpass filters. At present, the bandpass is set to 700--800~MHz, corresponding to redshift $1.03>z>0.78$ for the 21~cm line. A summary of the design parameters of the cylinder array is given in Table \ref{tab:cylinder_parameter}. 
The frequency response of the electronics in the signal channels is nearly flat, with a deviation of $\sim$1~dB across the 700-800~MHz band.

\section{Model}
\label{sec:model}

When radio signals are transmitted through a medium, a small fraction can be reflected at the interfaces. For example, incoming radio waves that are focused on the feed antennas may be partially reflected back to the cylinder surface and then reflected again to the feeds. The electric signals on the feed cables may be reflected by the connectors at the optical transmitter if their impedance is not perfectly matched with the cable. And this reflected signal may be reflected again at the other end of the cable. These reflections within the radio telescope system create ripples in the spectra of its outputs.  

If the voltage without reflection is $\mathcal{E}$, then in the presence of one reflection \citet{Kern:2019ytc} models the total signal as
\begin{equation}
  \mathcal{E}^\prime (\nu,t)= \mathcal{E}(\nu,t)+ \epsilon(\nu) \mathcal{E}(\nu,t),
\end{equation}
where $\epsilon (\nu)$ is the reflection coefficient of the signal at frequency $\nu$. It can be written in the form  
\begin{equation}
\epsilon (\nu) =  A
e^{i(2\pi \nu \tau + \phi)},
\end{equation}
where $\tau$ is the delay of the reflected signal and $\phi$ is the phase shift induced by the reflection. Here  
time and frequency stability of the reflection parameters are assumed, though these parameters will have some variations. The resulting auto-correlation visibility is
\begin{eqnarray}
V^\prime &=& [1+ (\epsilon + \epsilon^*) + \epsilon^* \epsilon)] \langle
\mathcal{E}^* \mathcal{E}\rangle \\
&=& [1+2A\cos(2\pi\nu\tau+\phi)+A^2] V,
\end{eqnarray}
where the $\langle \rangle$ in the first line denotes a short time average. 

More generally, if there are a number of interfaces, for which the signal in the following stage is 
\begin{equation}
  \mathcal{E}^{\prime}_i = (1+ \epsilon_i) \mathcal{E}_i,
\end{equation}
then the output voltage after multiple interfaces will be 
\begin{equation}
  \mathcal{E}^{\prime}(\nu,t) =\mathcal{E}(\nu,t)  \prod_{j} [1+\epsilon_j(\nu)],    
\end{equation}
where $\epsilon_j = A_j \exp[i(2\pi\nu\tau_i+\phi_i)]$.
Then the auto-correlation visibility is 
\begin{equation}
  V^\prime = V \prod_{i} [1+2A_i\cos(2\pi\nu\tau_i+\phi_i)+A_i^2].   
  \label{eq:one-reflection}
\end{equation}

In \citet{Kern:2019ytc}, only a single reflection is considered. This is a good approximation if the reflection amplitude is not large. However, if the reflection coefficient is stable, one may also consider multiple reflections at an interface. In each reflection the same factor is induced, so that 
\begin{equation}
  \mathcal{E}^\prime = \mathcal{E}(\nu,t) + \epsilon(\nu) \mathcal{E}(\nu,t) 
  + \epsilon^2 \mathcal{E} + ...  
  = \frac{1}{1-\epsilon}\mathcal{E}(\nu,t)
\end{equation}
and 
\begin{equation}
  V^\prime(\nu,t) = \frac{1}{1-2A\cos(2\pi\nu\tau+\phi)+A^2} V(\nu,t).
\end{equation}
This can also be generalized easily to the case of multiple reflecting interfaces. The voltage in this case is given by 
\begin{equation}
  \mathcal{E}^\prime = \mathcal{E} \prod_{j} \frac{1}{1-\epsilon_j}    
\end{equation}
and the auto-correlation is
\begin{equation}
  V^\prime(\nu,t) = V(\nu,t)  \prod_{i} \left[\frac{1}{1 - 2 A_i\cos(2\pi\nu\tau_i + \phi_i) + A_i^2}\right]. 
  \label{eq:multi-reflection}
\end{equation}

The delay, $\tau$, is the time required for a round trip between the two reflecting interfaces in the signal chain and is given by
\begin{equation}
  \tau = \frac{2L}{v},
\end{equation}
where $v$ is the wave speed and $L$ is the distance between the two reflecting interfaces, i.e. the medium length. For the open space between the feed and reflector, $v \approx c$, while for the coaxial cable $v = c / \sqrt{\varepsilon_r \mu_r }$, where  $\varepsilon_r$ and $\mu_r$ are, respectively, the relative electric permittivity and magnetic permeability of the dielectric material. The wave speed in the cable is approximately $0.7c$ for commonly used coaxial cables \citep{pozar2009microwave}.

Therefore, the reflections may induce sinusoidal oscillations in the spectrum. These will cause distortions in the correlation amplitude. By analyzing possible standing wave structures in the bandpass, one can determine where the standing waves are produced and mitigate the impedance mismatch problem. 

Besides the astronomical signals, there is also receiver noise. For the auto-correlation, the receiver noise background is usually much stronger than the signal from astronomical sources. The noise generated in the preceding stages of the receiver system is amplified in the same signal chain and is subject to the same reflection effects. Indeed, at some resonant frequencies, the reflected signals may have the same phases as the original ones such that stable standing waves are generated.

\section{Analysis}
\label{sec:analysis}

In this work, we use the data taken over 5 days from 2016/09/27 to 2016/10/02 and 9 days from 2018/03/22 to 2018/03/31 for analysis.  

\begin{figure}[htbp]
  \centering
  \includegraphics[width=0.49\textwidth]{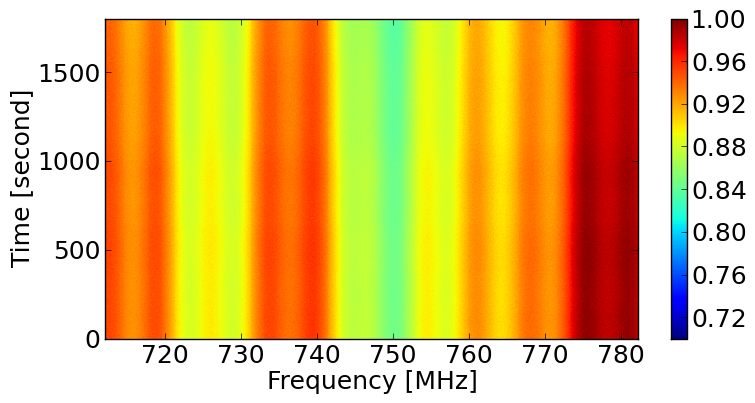}
  \includegraphics[width=0.41\textwidth]{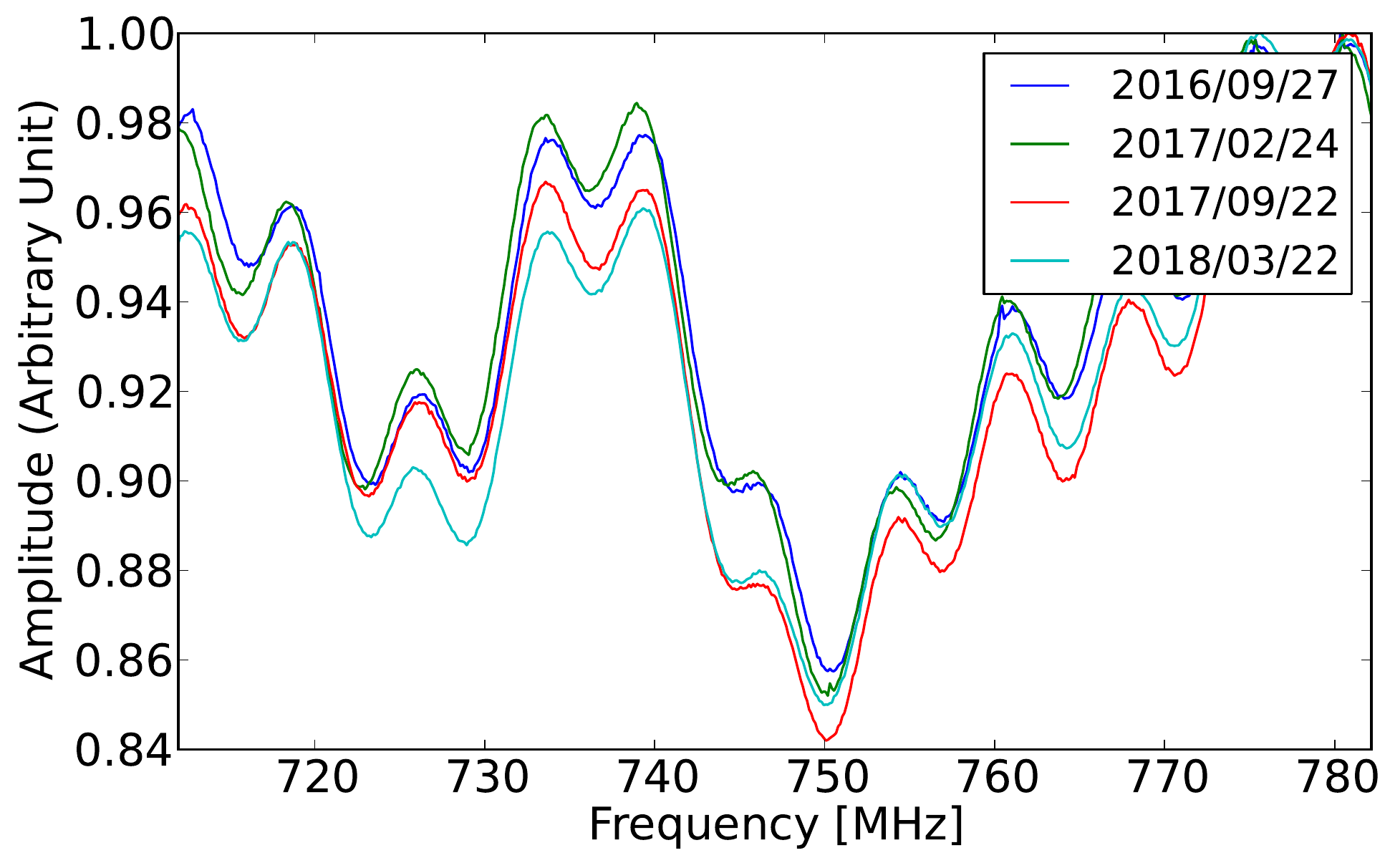} \\
  \includegraphics[width=0.49\textwidth]{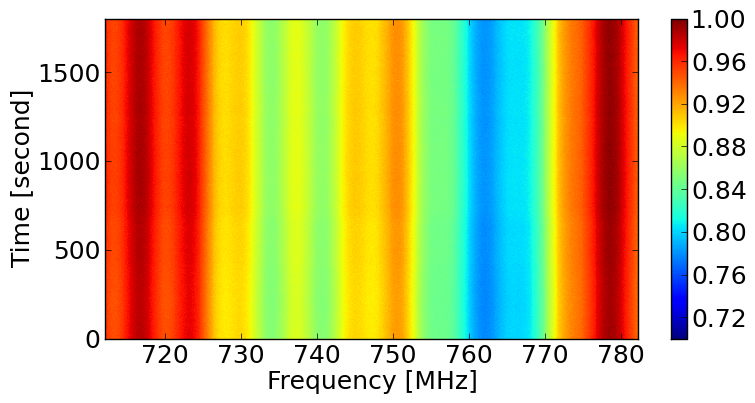}  
  \includegraphics[width=0.41\textwidth]{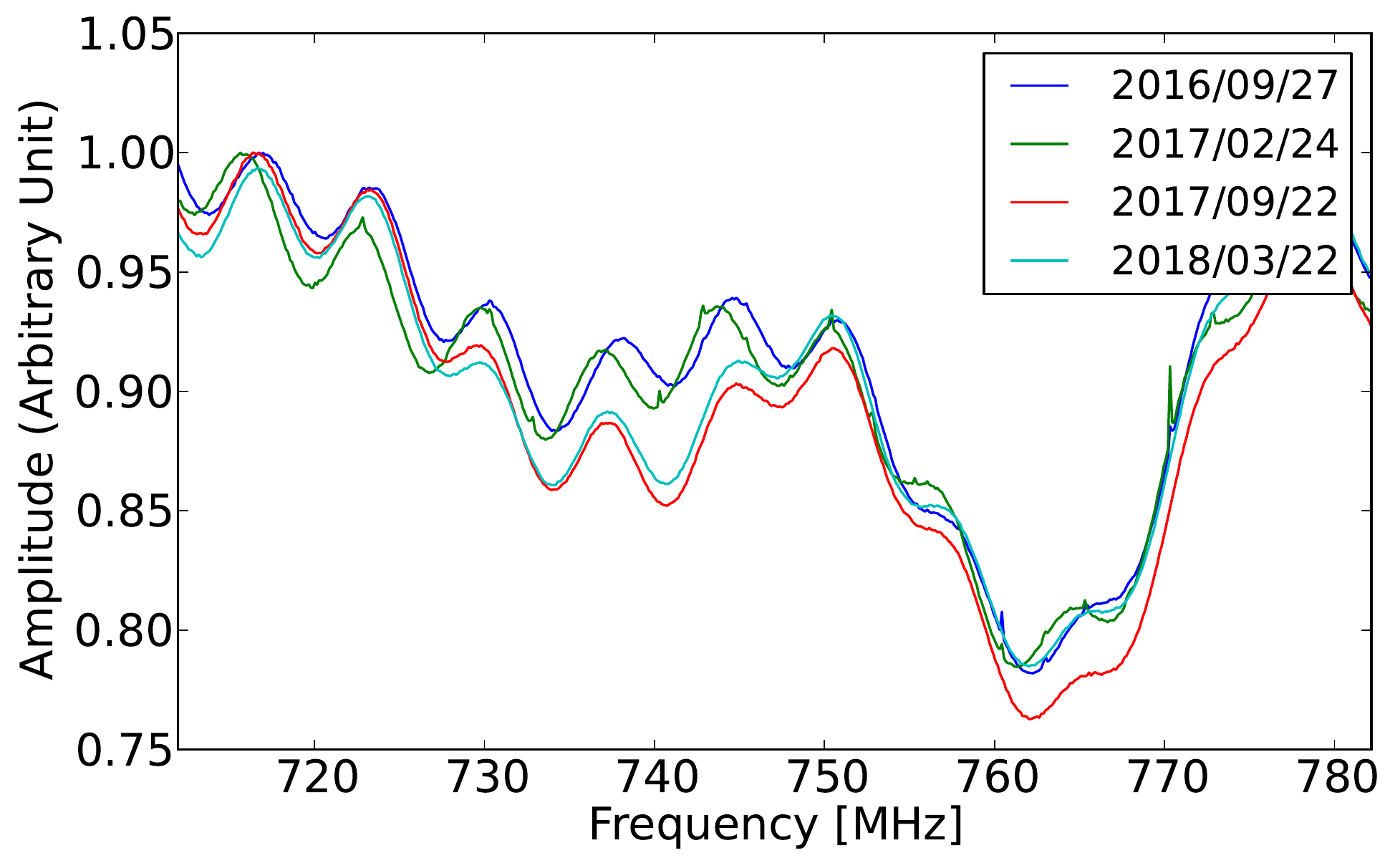} \\
  \caption{Left: amplitudes of auto-correlations B22X (top) and B22Y (bottom) over 30 minutes. Right: amplitudes of auto-correlations B22X (top) and B22Y (bottom) on different days. }
  \label{fig:band_stable_time}
\end{figure}

In Fig.~\ref{fig:band_stable_time} we show two typical auto-correlation visibilities. In the left panel, we show the auto-correlation in a ``waterfall plot'', i.e. as a function of frequency (abscissa) and time (ordinate) during 30 minutes of observation at night. The observation time is chosen such that there is no strong radio source transiting through the beam, and we shall refer to this case as {\it blank sky}, though of course at any time there are many weak sources in the field of view. In the right panel, we plot the same auto-correlations taken on different days. We can see that there are frequency structures which are very stable over months and years. 

\begin{figure}[htbp]
  \centering
  \includegraphics[width=0.4\textwidth]{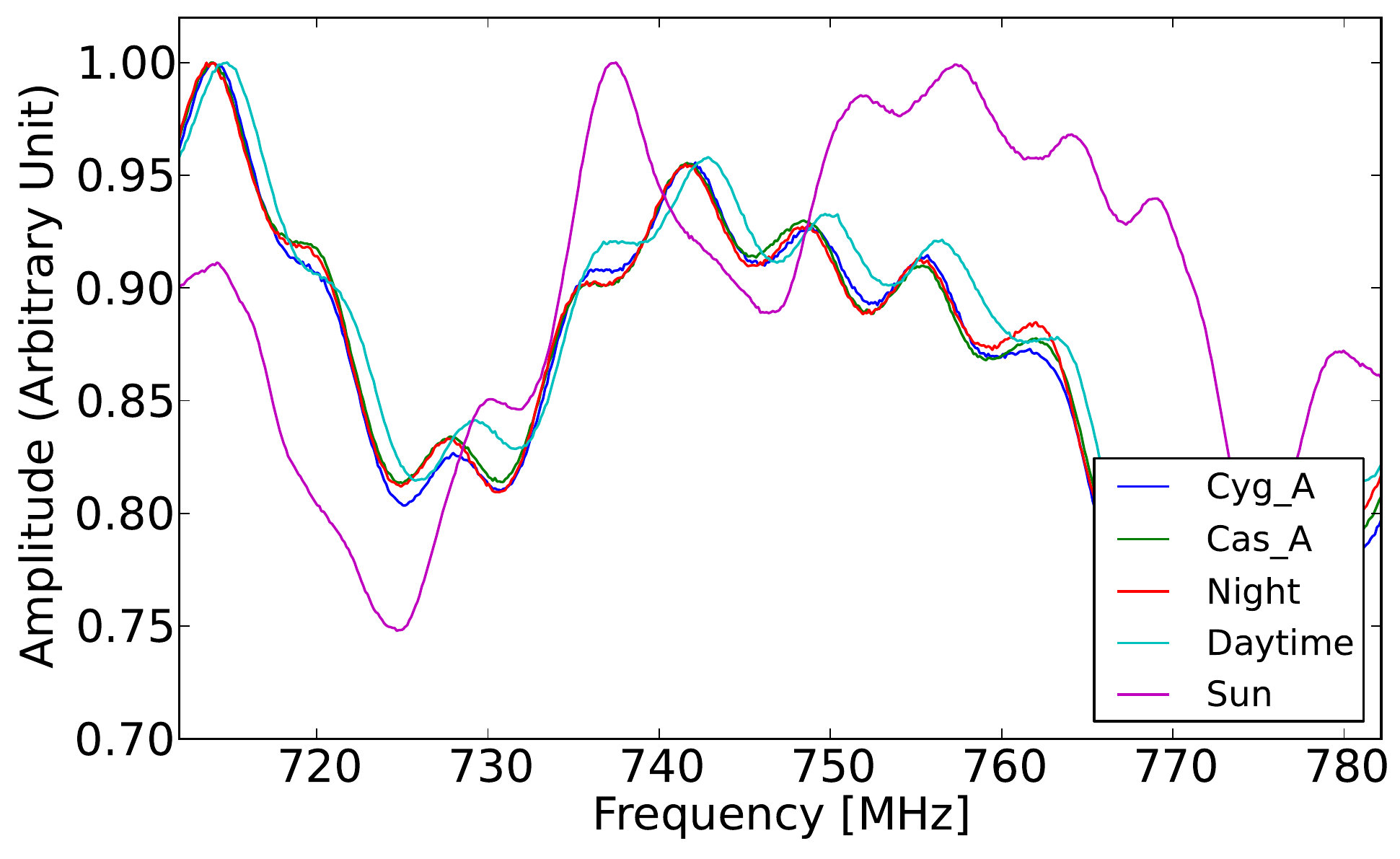}
   \includegraphics[width=0.4\textwidth]{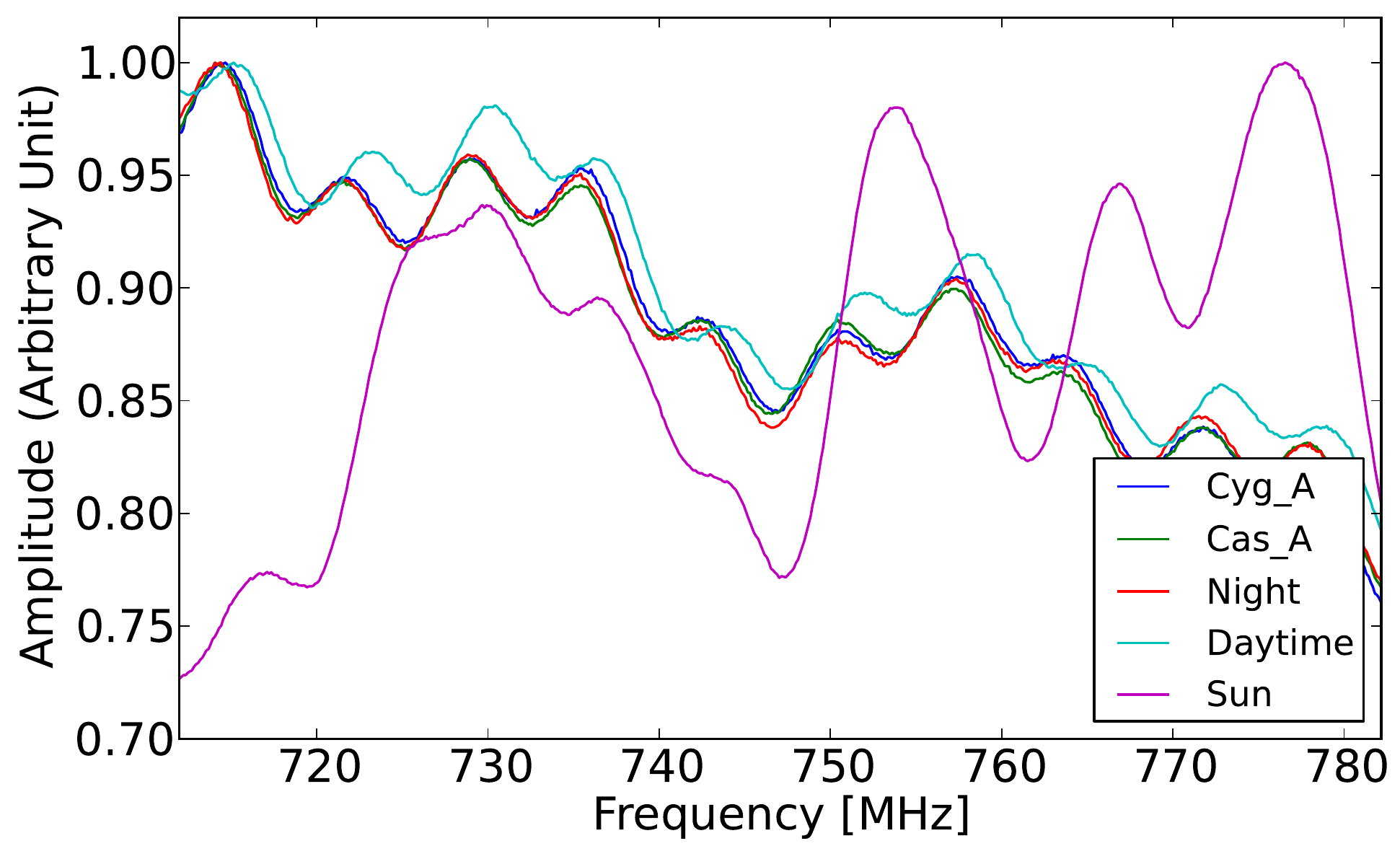}\\
     \includegraphics[width=0.4\textwidth]{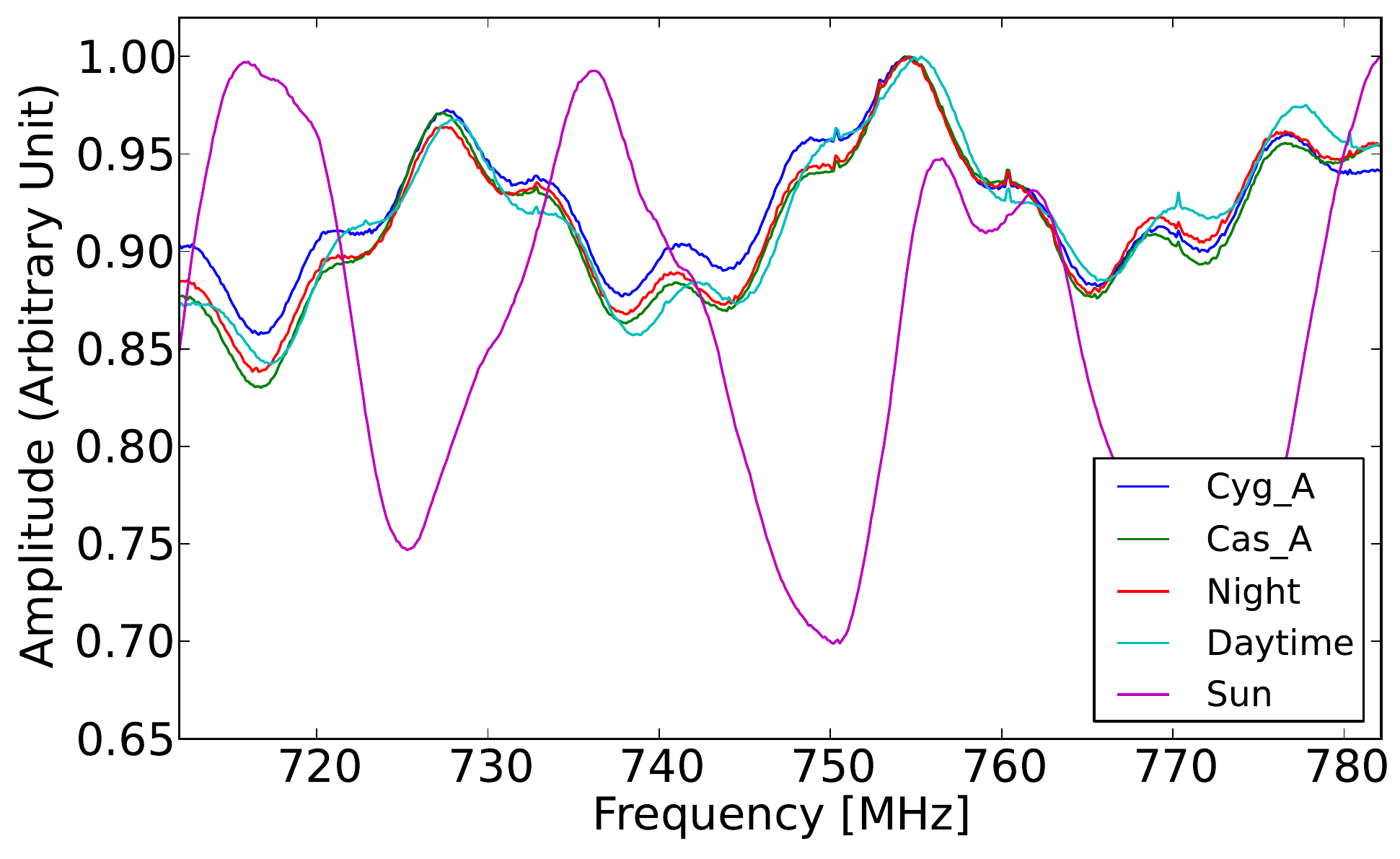}
   \includegraphics[width=0.4\textwidth]{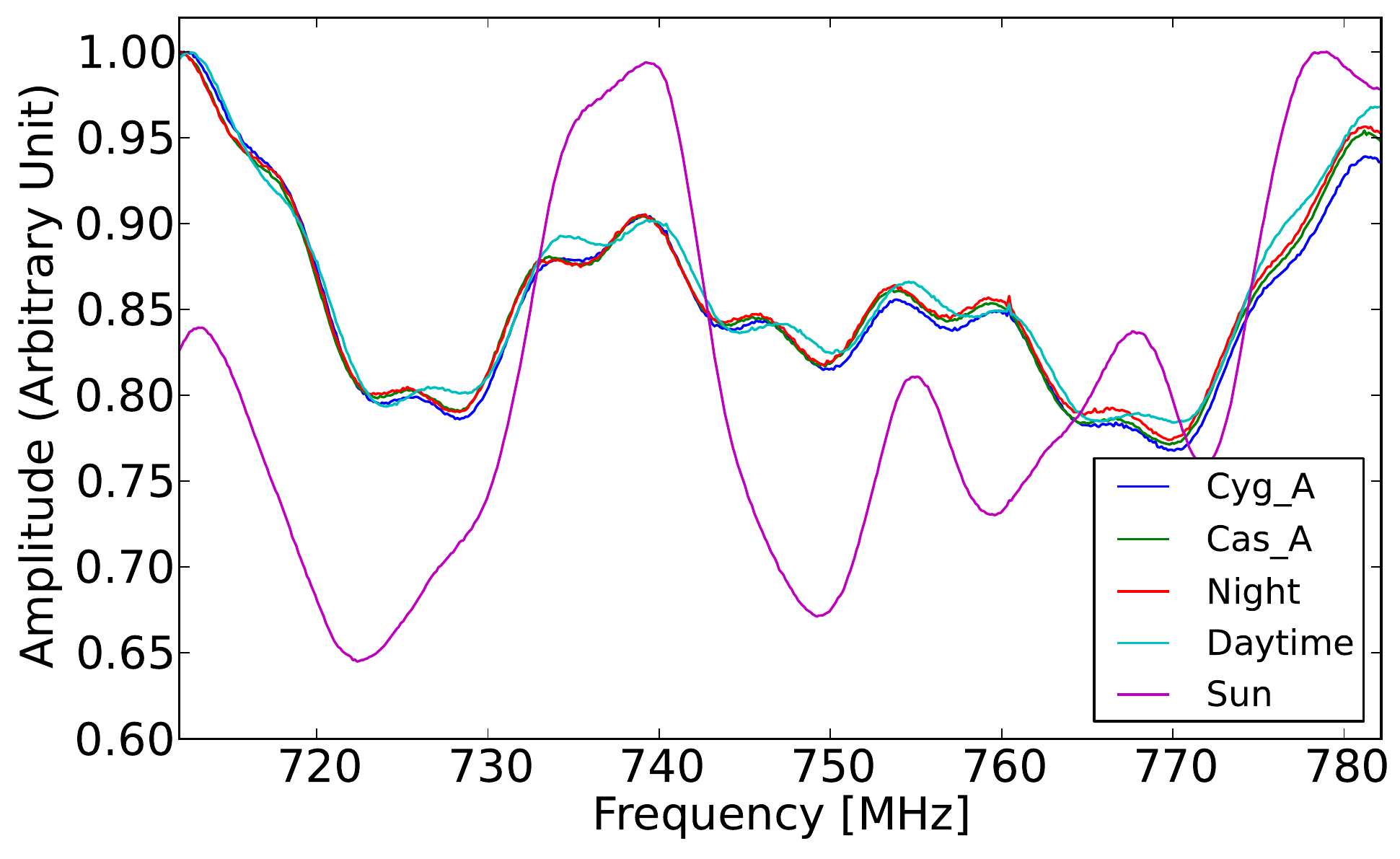}\\
  \caption{The amplitudes of auto-correlation visibilities for four typical signal channels (Top left: A1Y; Top right: B15Y; Bottom left: A3X; Bottom right: A23X), obtained by averaging over 40 seconds during the transit of Cyg~A (blue), Cas~A (green), blank sky at night (red), daytime (cyan), and the Sun (magenta).}
  \label{fig:band_stable_src}
\end{figure}

With the understanding that these frequency structures are stable over time for the blank sky, in Fig. \ref{fig:band_stable_src}, we plot the spectra of some typical signal channels when different radio sources transit, as well as daytime and nighttime blank sky. For each signal channel, the detailed structure is different, but we can see that they all appear to have some oscillatory modulations. The auto-correlations of different receiver channels have different spectra, but for each of them, the spectrum is almost the same when the telescope is observing the blank sky, i.e. the part of sky devoid of strong radio sources, or observing most bright radio sources. The only exception is the Sun; during its transit the frequency structure becomes quite different. This may be understood as follows: the auto-correlation is largely dominated by the stable receiver noise, even when observing strong radio sources such as Cyg~A or Cas~A. However, the Sun is very bright, which induces a signal with strength comparable to the receiver noise. The Sun also has a large zenith angle (ZA) of $\sim 46^\circ$ during the observations studied here, so the induced standing wave patterns on the cylinder antenna may also differ from those present during the blank sky time.

Below, we study the standing waves for the case of the blank sky and the radio source (after subtracting off the background) separately. We shall use the Fourier transform as a tool. The Fourier transform of the frequency spectrum is usually called the {\it delay transform}, and the result is called a {\it delay spectrum}, with the frequency $\nu$ and delay $\tau$ forming a conjugate pair:
\begin{equation}
  \tilde{V}(\tau) = \int W(\nu) V(\nu) e^{i2\pi \nu\tau} \mathrm{d}\nu,
\end{equation}
where $W(\nu)$ is the window function.
In discrete form,
\begin{equation}
\tilde{V}(\tau) = \sum_{n} W_n V(\nu_n) e^{i2\pi \nu_n \tau}.    
\end{equation}
As the auto-correlation is a real number, the delay spectrum is always symmetric in our case. Below we shall only plot the positive frequency part.

The observed spectrum has 576 frequency channels with a resolution of 122 kHz. We use a 16384-point fast Fourier transform (FFT) with zeros padded in the end of the spectrum to obtain a denser sampling in Fourier space.  The frequency spectrum is multiplied by a window function before the FFT to reduce spectral leakage. Here we used the Hann window function,
\begin{equation}
  W(n) = 0.5 - 0.5 \cos{\left( \frac{2\pi n}{M - 1}\right)}, \qquad(0 \leq n \leq (M - 1)),
\end{equation}
where $M = 576$ is the window length. 
Trials with a few other window functions and window lengths show that there are minor differences in the result, but will not affect the conclusions drawn below.

\subsection{Blank Sky}

\begin{figure}[htbp]
  \centering
  \includegraphics[width=0.48\textwidth]{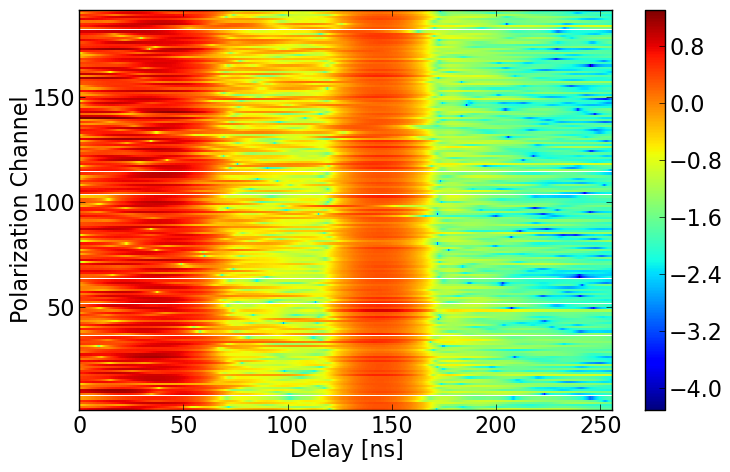}
  \includegraphics[width=0.48\textwidth]{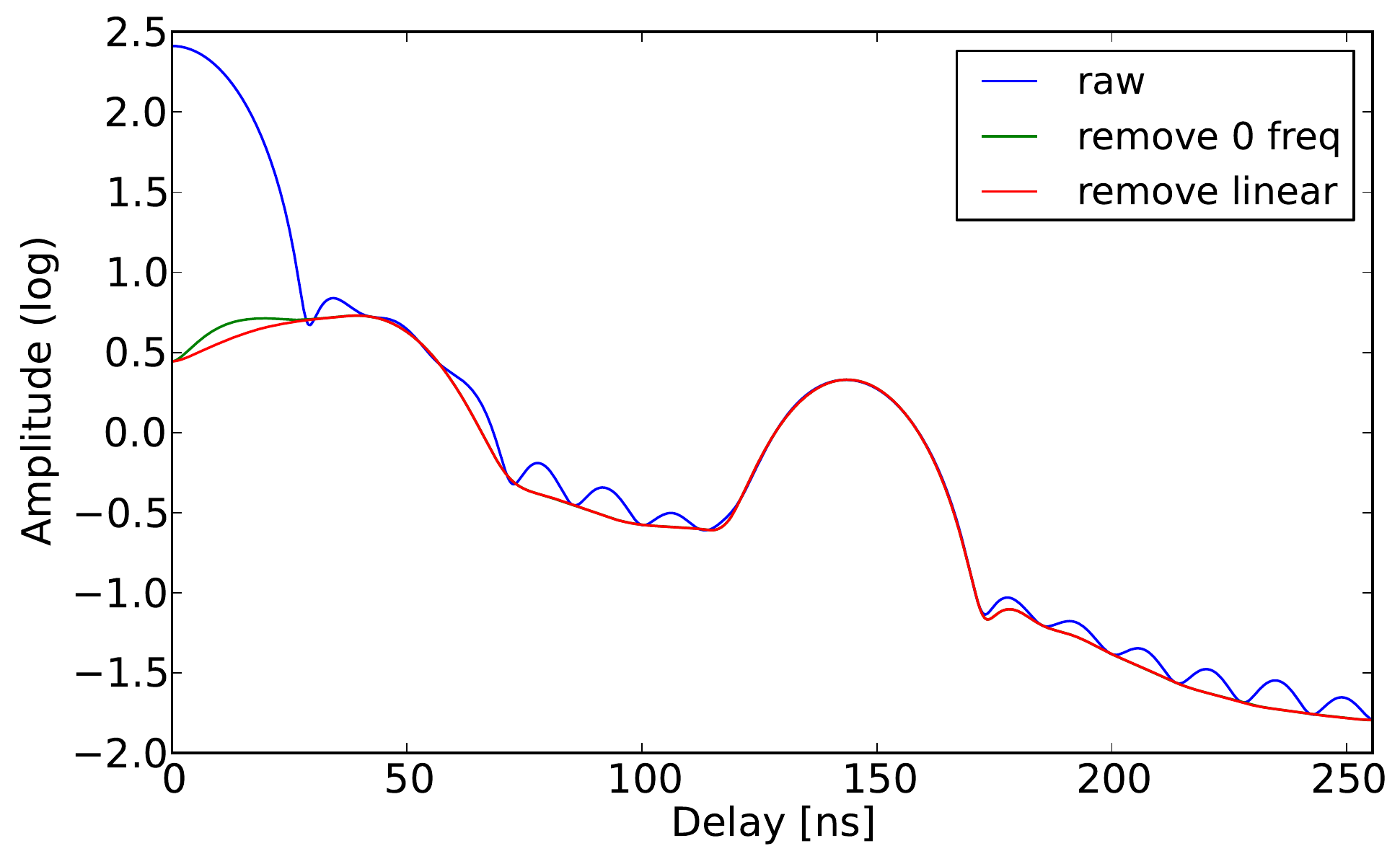}
  \caption{The delay transform for nighttime blank sky. Left panel: the delay spectra for the 192 signal channels. Right panel: The  delay spectra of the average of all 192 channels. 
  }
  \label{fft2d_night}
\end{figure}

The delay transform results are shown in Fig. \ref{fft2d_night}. In the left panel, the delay spectra of all of the 192 signal channels are plotted, arranged along the vertical axis. In the right panel, the average spectrum is shown. Three curves are plotted: the blue curve marked as ``raw'' is the original spectrum; the green curve marked as ``remove 0 freq'' is obtained by first subtracting the mean of the spectrum in the observed frequency range before FFT. This operation removes the large zero frequency component, so that smaller Fourier components can be seen more clearly. The red curve marked as ``remove linear'' is  obtained by further subtracting the linear component in the frequency, i.e. the overall ascending or descending trend in the observed frequency range.  

As expected, for most channels the spectra are very similar to each other. There are a few channels which differ from the normal, as shown very clearly as white lines. These are malfunctioning channels and have been masked during the subsequent analysis. For the majority of channels, the most prominent feature is a high plateau or broad peak from 0 to $\sim 60 \ns$, and also a peak at around 142~ns. This structure can be seen more clearly in the average spectrum of all channels, as shown in the right panel of Fig.\ref{fft2d_night}. 

There is a clear peak at delay time of $\sim 142 \ns$, corresponding to 29.8~m assuming a standard wave velocity of $0.7c$, which is very close to the double length of the 15~m feed cable. If we assume that it is indeed induced by the 15 m feed cable, we obtain a wave velocity of $0.7047c$, which is very close to the value we measured for the cable. 

The spectrum is also high between 0 to 60 ns. If we remove the zero frequency component by subtracting the average, there is still a broad peak, and for some feeds two blended peaks can be seen. 

\begin{figure}[htbp]
    \centering
    \includegraphics[width=0.45\textwidth]{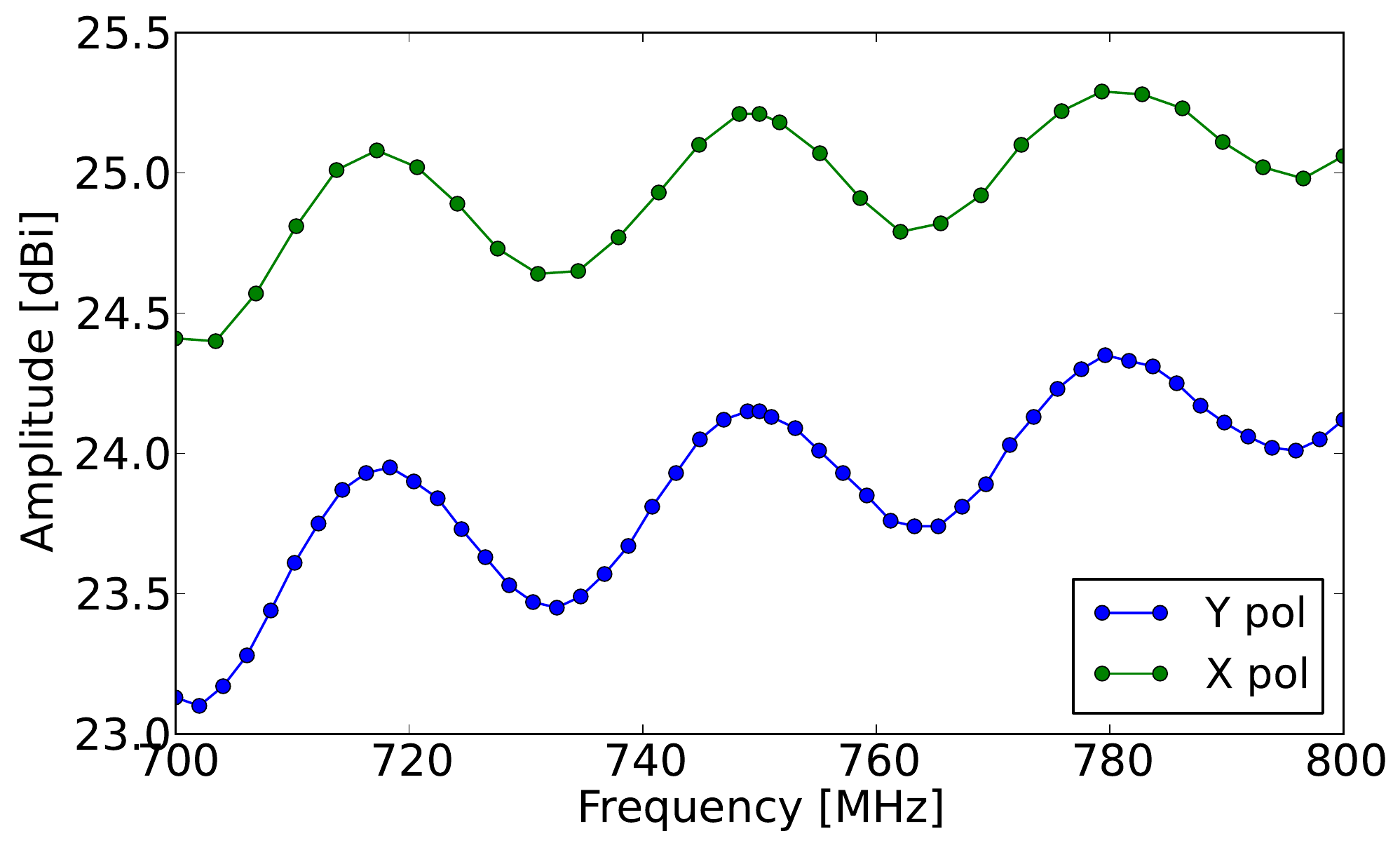}
    \includegraphics[width=0.45\textwidth]{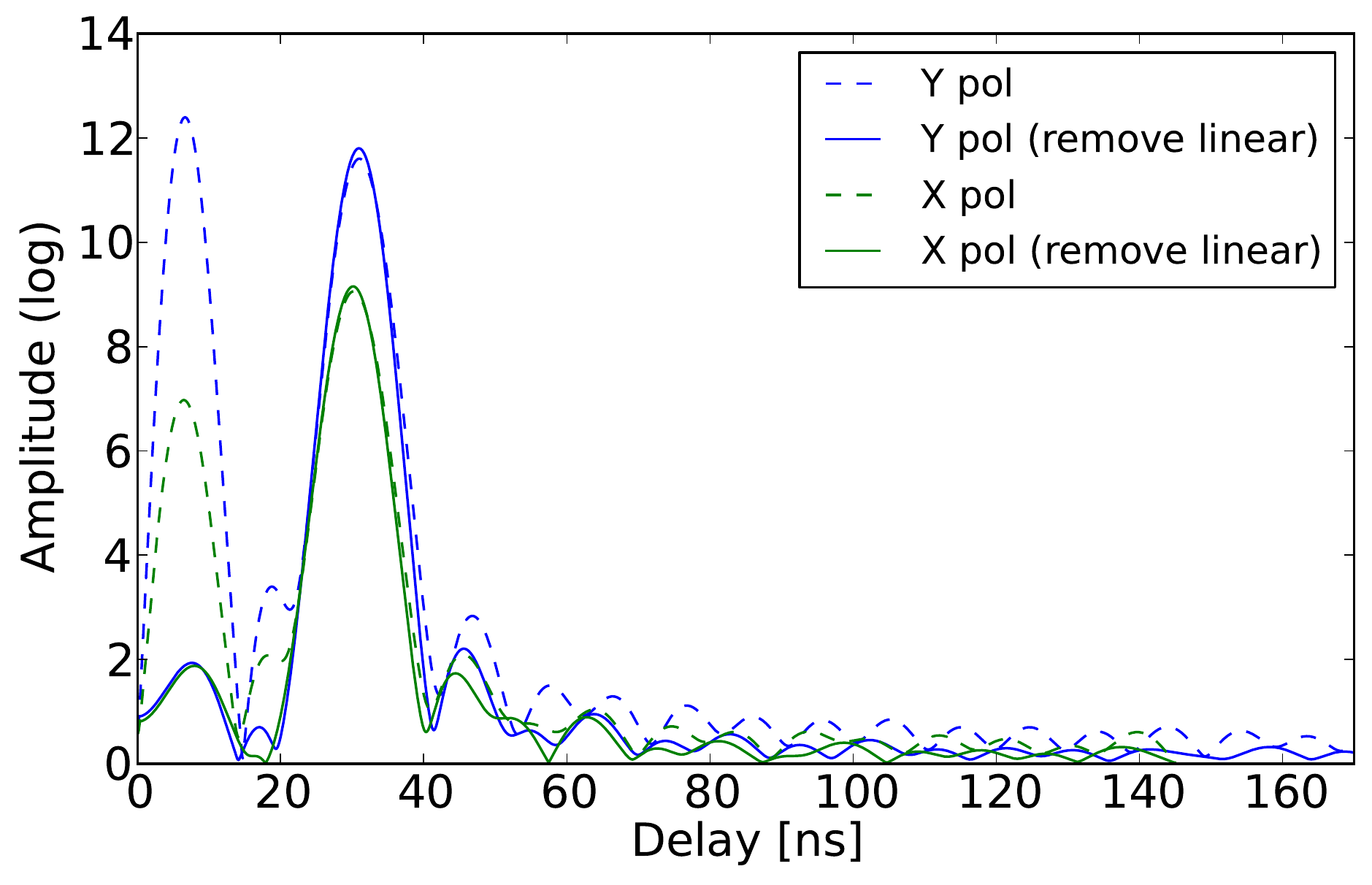}
    \caption{Left: the simulated directivity of the cylinder with one feed. Right: the corresponding delay transform. }
    \label{fig:sim_band}
\end{figure}

To better understand the origin of these peaks, we look at the simulated center (zenith) directivity of the cylinder antenna as a function of frequency. The directivity is defined as 
\begin{equation}
  D = \frac{P_{\rm max}(\theta, \phi)}{P_{\rm av}},
\end{equation}
where $P_{\rm max}(\theta, \phi)$ is the maximum power of the antenna, while the average power is 
\begin{equation}
P_{\rm av}=\frac{1}{4\pi}\int P(\theta,\phi) \mathrm{d}\Omega
\end{equation}
so that $D=4\pi/\Omega_A$. The simulation is performed with the \texttt{CST Studio} software, with one feed placed at the center of the cylinder focus line. The simulation result is shown in the left panel of Fig. \ref{fig:sim_band}.  We can see a clear sinusoidal modulation over the frequencies in the simulated center directivity. A delay transform of this is shown in the right panel; two prominent peaks can be seen in the spectrum. The first one is at 7~ns and the second is at 31~ns. The second peak corresponds to a distance of 9.3~m, nearly twice the focal length (4.8 m), indicating that the modulation may be associated with standing waves on the cylinder antenna. The physical origin of the first peak is less clear, but it is related to the ascending bandpass curve. If we remove this overall ascending trend by subtracting the linear fit to the curve, the 7~ns peak disappears, as shown in Fig. \ref{fig:sim_band}.

This simulation is not completely accurate in representing the features of the cylinder, as we included only one feed, while in reality there are multiple feeds that will affect each other, and many small details of the antenna (e.g. the supporting struts, cables, and surface error) can not be completely represented. However, this exercise does show that a part of the low frequency broad peak seen in the observation data may be due to the overall trend of the spectrum in this band, and removing the overall trend could help reveal the other contributors.

Inspired by this test with simulation data, we also remove the overall ascending or descending trend within the band in the observational data. In the delay spectrum, there is still a broad peak around 35 ns. This, as also shown in the case of the simulation, is fairly close to the delay length of twice the focal length of the reflector (32 ns). 

However, if the peak is induced by standing waves, another possible location of the standing wave is the IF cable. As described in Sec. \ref{sec:instrument}, it is actually made up of two cables connected via SMA connectors. If we view them as separate cables with wave velocity of $0.7c$, the delay is 19~ns, while if we regard them as a whole it is 38~ns. The standing waves may exist both between the feed and reflector, and within the IF cables, and they may all mix together to form the low frequency plateau or broad peak seen in the delay spectrum. 

\subsection{Radio Sources}
\label{sec:radio_sources}

Below we consider the spectrum during the transit of radio astronomical sources, which may differ from the background noise. To extract the signal induced by the source, we look at the difference between the peak of the radio source transit and the blank sky background before or after the transit.

We studied the transit of several bright astronomical sources, including Cyg~A, Cas~A and the Sun. The radiation spectrum of Cyg~A and Cas~A is modelled as a polynomial expansion in the frequency range of interest,
\begin{equation}
  \log{S} = \sum_{n=0}^5 a_n [\log(\nu)]^n
\end{equation}
where for Cyg~A $a_0 = 3.3498, a_1 = -1.0022, a_2 = -0.2246, a_3 = 0.0227, a_4 = 0.0425$ and for Cas A $a_0 = 3.3584, a_1 = -0.7518, a_3 = -0.0347, a_4 = -0.0705$ \citep{2017ApJS..230....7P}. In our observation band (700--800~MHz), the flux difference between the low and high frequency side is about 5\%. The radio source spectrum results in a descending curve in the bandpass and can be calibrated by the polynomial spectrum.

\begin{figure}[htbp]
    \centering
    \includegraphics[width=0.45\textwidth]{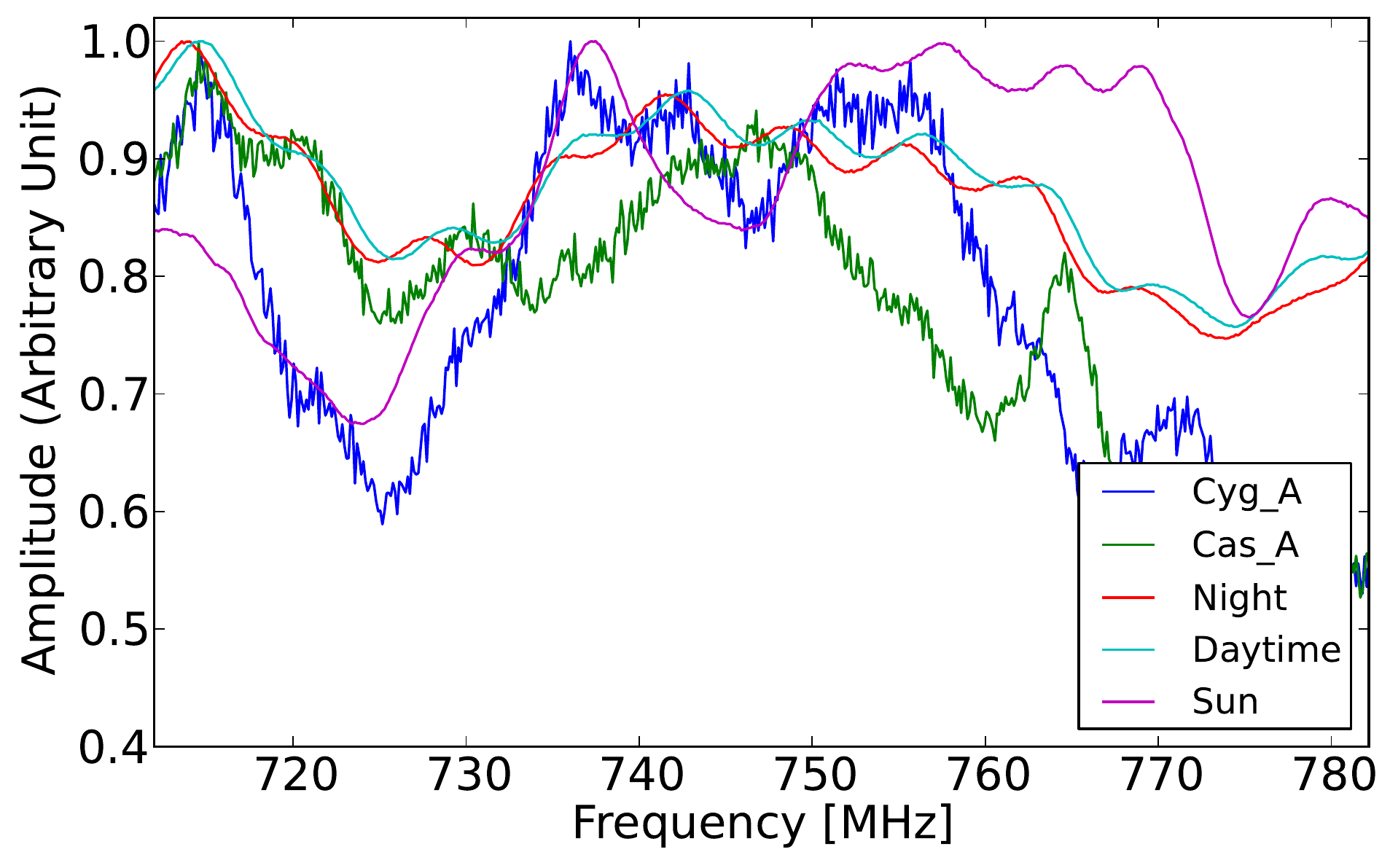}
    \includegraphics[width=0.45\textwidth]{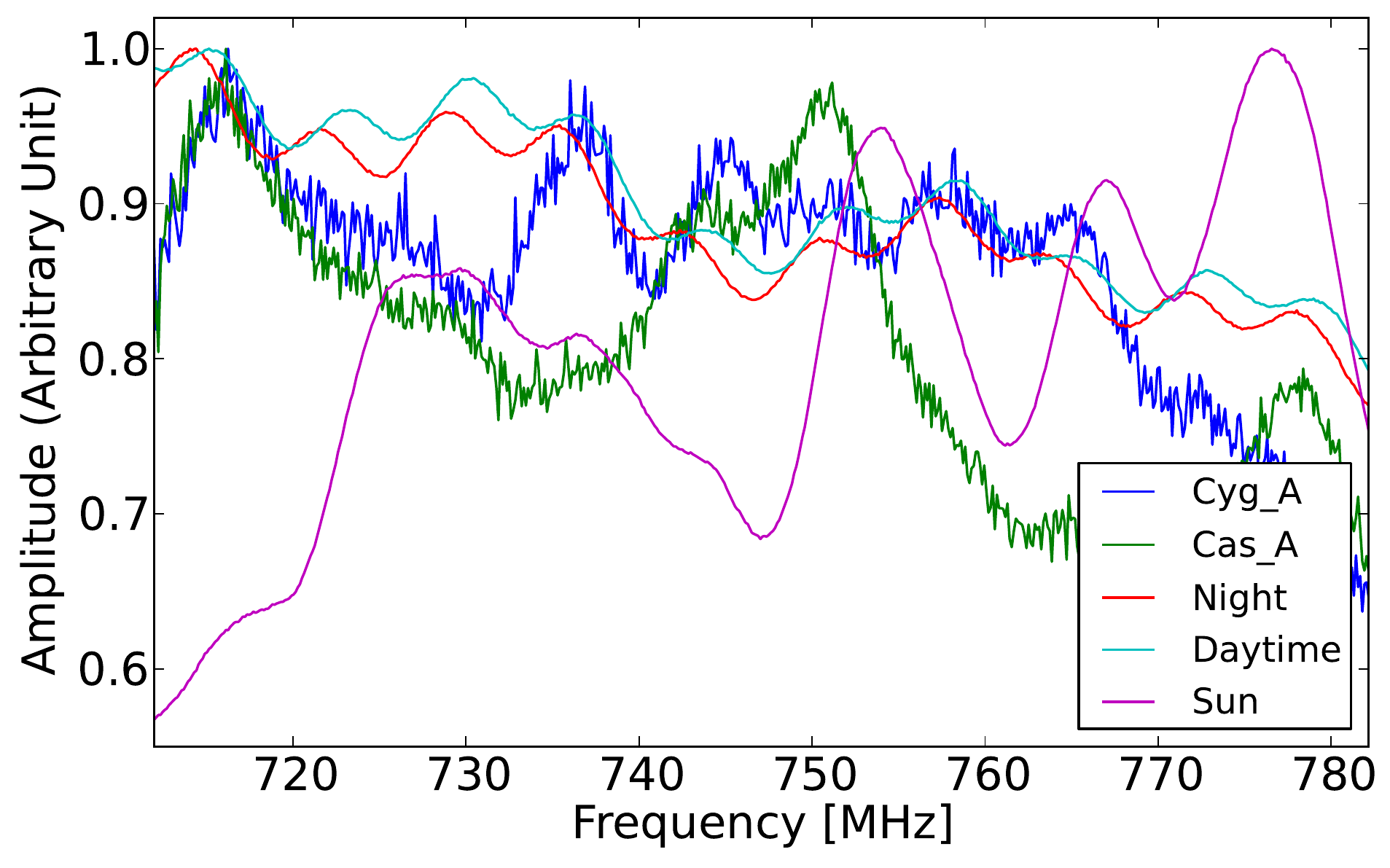}\\
    \includegraphics[width=0.45\textwidth]{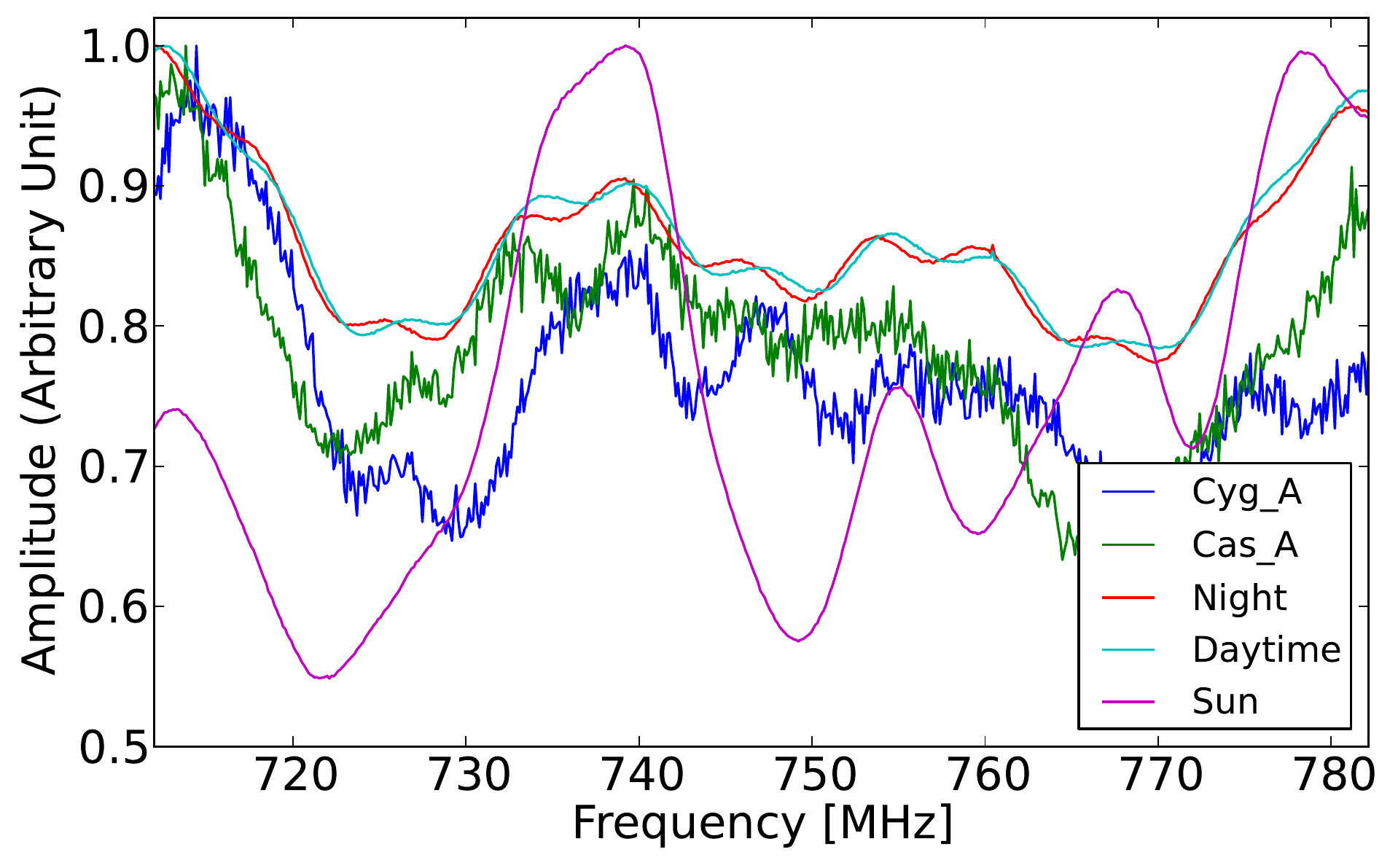}
    \includegraphics[width=0.45\textwidth]{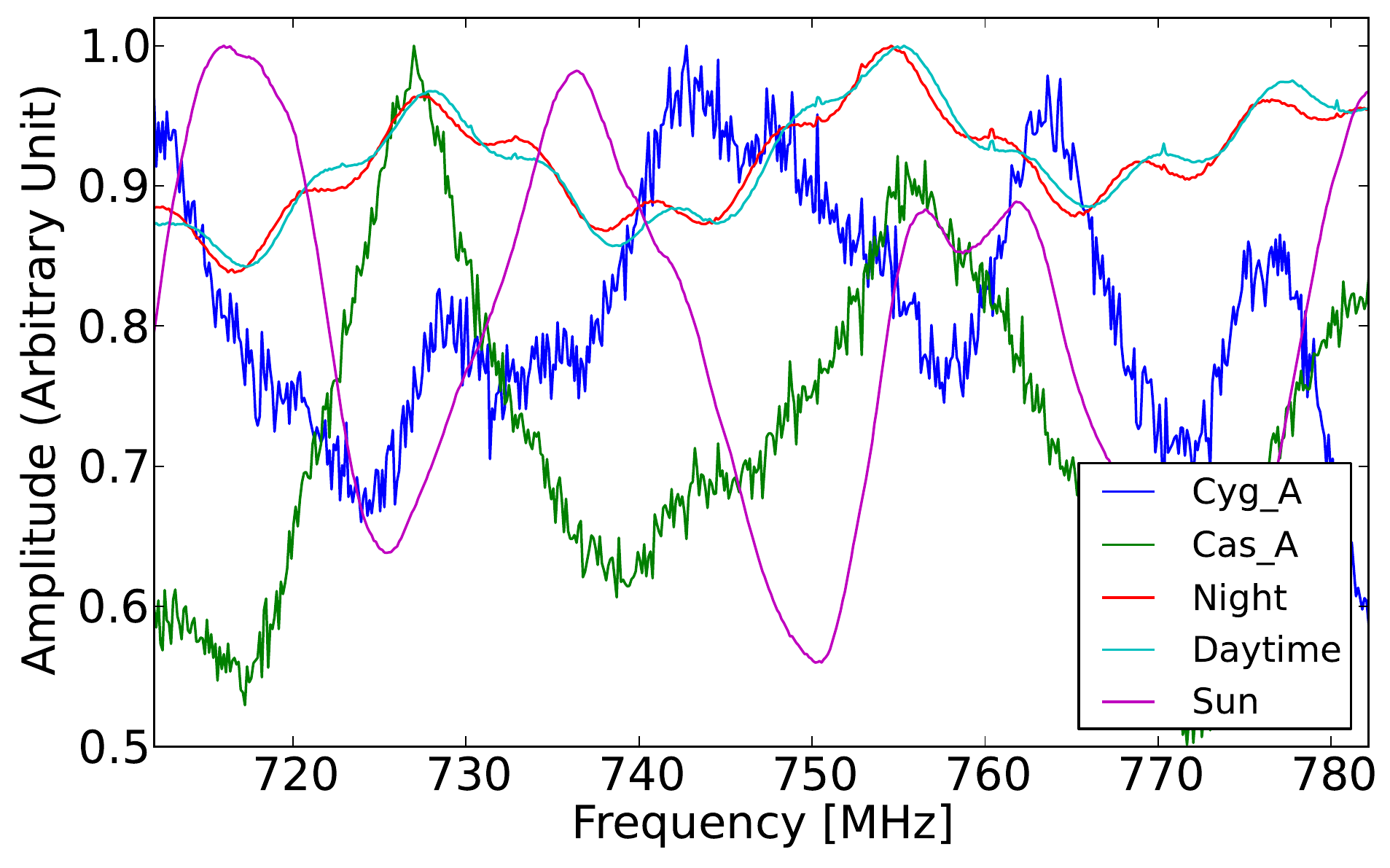}\\
    \caption{Spectra of 4 typical auto-correlations with background noise removed. Top left: A1Y. Top right: A3X. Bottom left: A23X. Bottom right: B15Y.}
    \label{fig:bandcmpr_src_net}
\end{figure}

The spectra during the transit of Cyg~A, Cas~A, and the Sun, as well as the background during nighttime and daytime for 4 typical feeds, are shown in Fig.~\ref{fig:bandcmpr_src_net}. Here, we have normalized each curve by its maximum. As we can see, the frequency response is somewhat different in each case. The nighttime and daytime backgrounds are actually quite close to each other. The Cyg~A, Cas~A, and Sun curves are all different. Such differences may originate partly from the different spectra of the sources, which are coupled with the non-flatness of the system frequency response, producing the different patterns in the result. But perhaps more important is that the zenith angles of these sources are very different from each other, so the reflections and standing waves on the cylinder antenna are different.

\begin{figure}[htbp]
  \centering
  \includegraphics[width=0.47\textwidth]{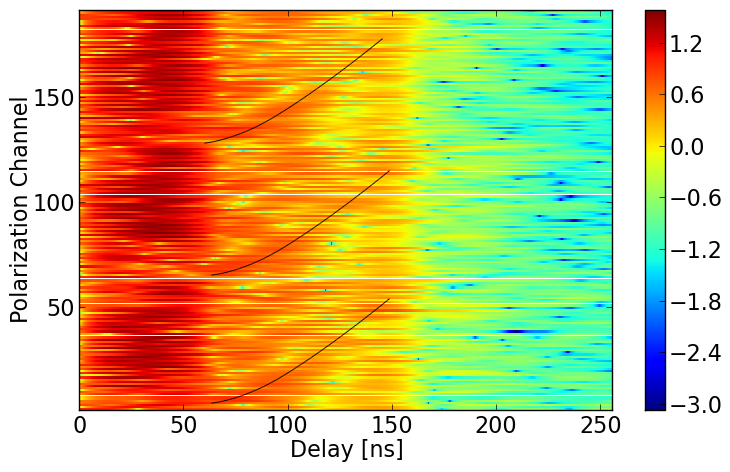}
   \includegraphics[width=0.47\textwidth]{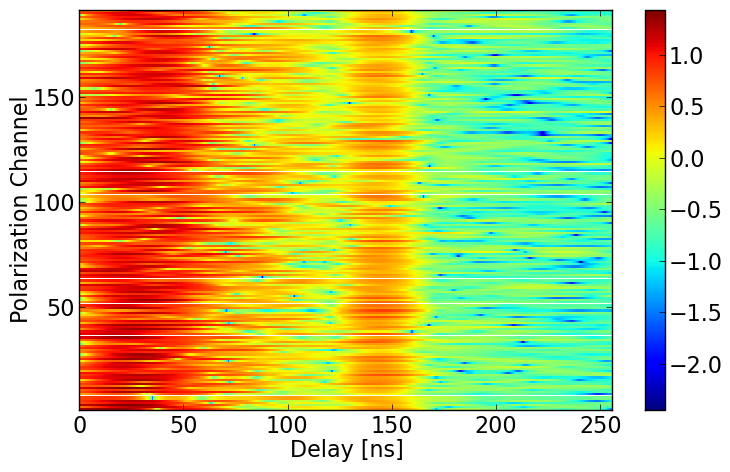}\\
 \includegraphics[width=0.47\textwidth]{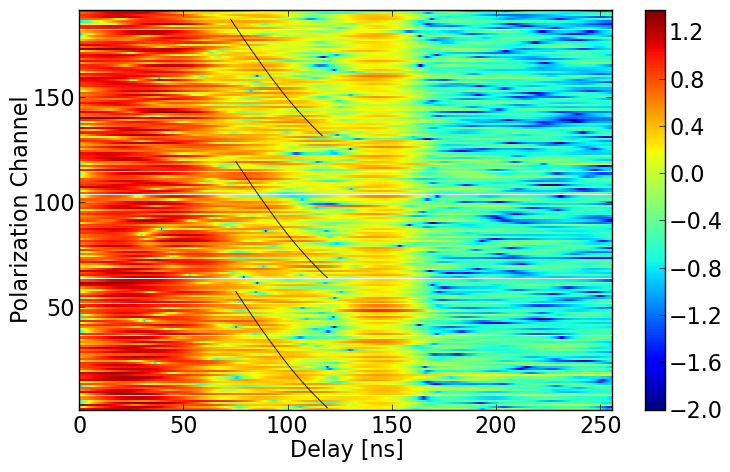}
 \includegraphics[width=0.47\textwidth]{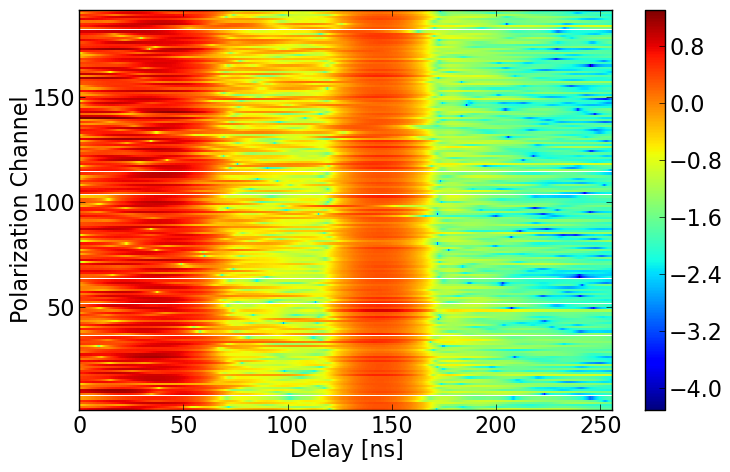}
  \caption{Time delays of all auto-correlations when different radio sources transit. Top left: Night. Top right: Cyg~A. Bottom left: Cas~A. Bottom right: Sun. The feather-like features discussed in the text are marked by black curves in the figure for Cas~A and the Sun.}
  \label{fig:all_fft_cmpr_src}
\end{figure}

The delay spectra for all the signal channels during the transit of the Cyg~A (top right), Cas~A (bottom left) and Sun are shown in Fig.~\ref{fig:all_fft_cmpr_src}. For comparison, we also show the delay spectra for blank sky during nighttime on top left. Compared with the background (blank sky), in these delay spectra of radio sources, the low frequency peaks spread out to higher values at 
$50 \ns <\tau<120 \ns$. For auto-correlation visibilities of the Tianlai array, the background is dominated by the noise within the electronic circuits, and much of the standing wave may be generated within the circuits. For the radio sources, however, the standing wave on the antenna itself becomes more prominent, which may explain the spread. 
Interestingly, we can see three ``feather-like'' structures in Cas~A and the Sun delay spectra on each of the antennas, with the Cas~A curving downward and the Sun curving upward. This is not seen in the Cyg~A case. These features indicate some sort of standing wave patterns along the north-south direction on the cylinder antennas, because the electronic circuits are independent and well isolated should not produce such features. Indeed, we do not see these features in the Cyg~A transit, which passes near the Zenith (ZA = $3.5^\circ$). The Sun passes to the south during the transit (ZA = $46^\circ$), while Cas~A passes to the north during the transit (ZA = $15^\circ$), which may be the reason why the ``feather'' bends ``upward'' and ``downward'' with different ratios in these cases, respectively.

\begin{figure}[htbp]
    \centering
    \includegraphics[width=0.47\textwidth]{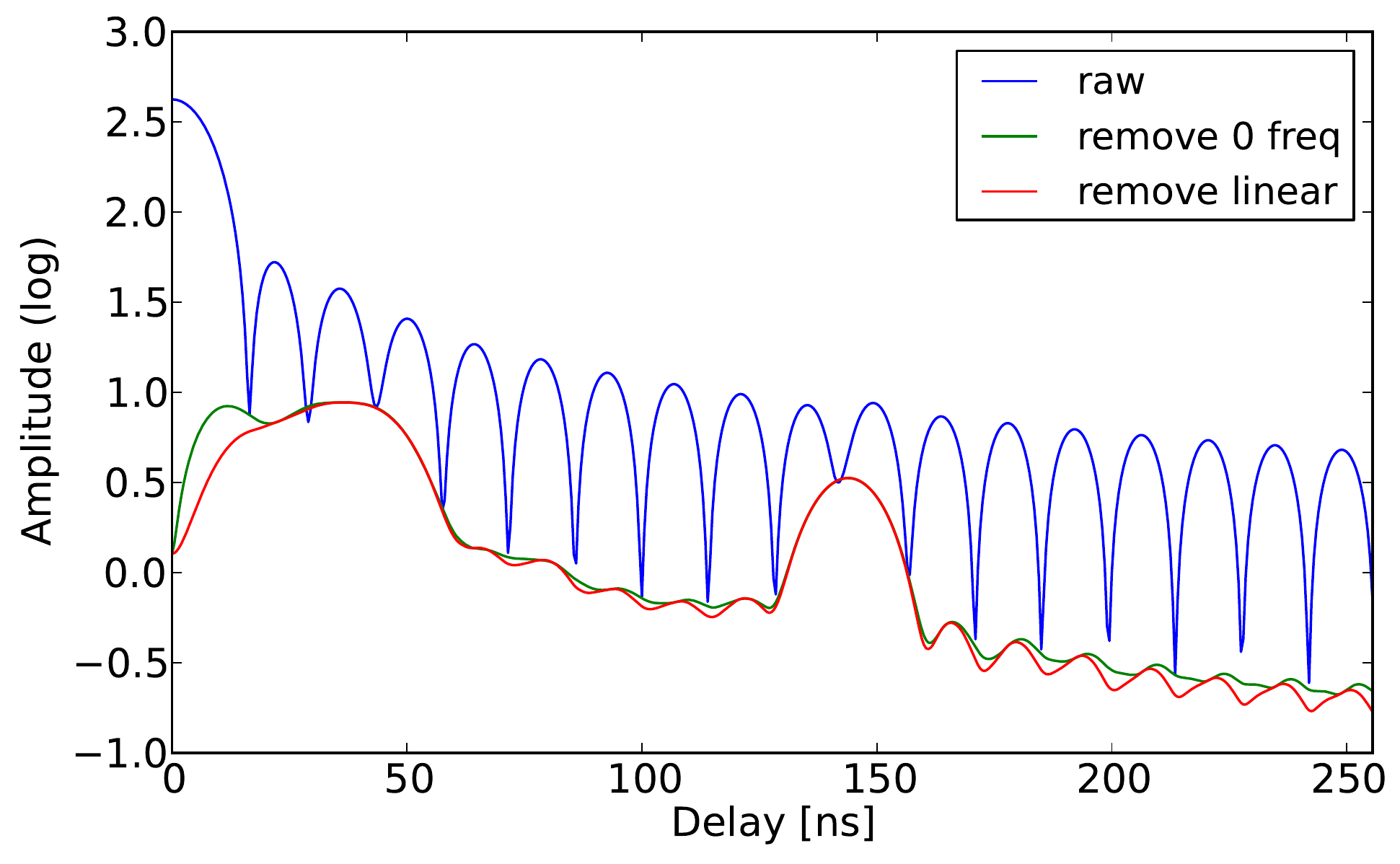}    
    \includegraphics[width=0.47\textwidth]{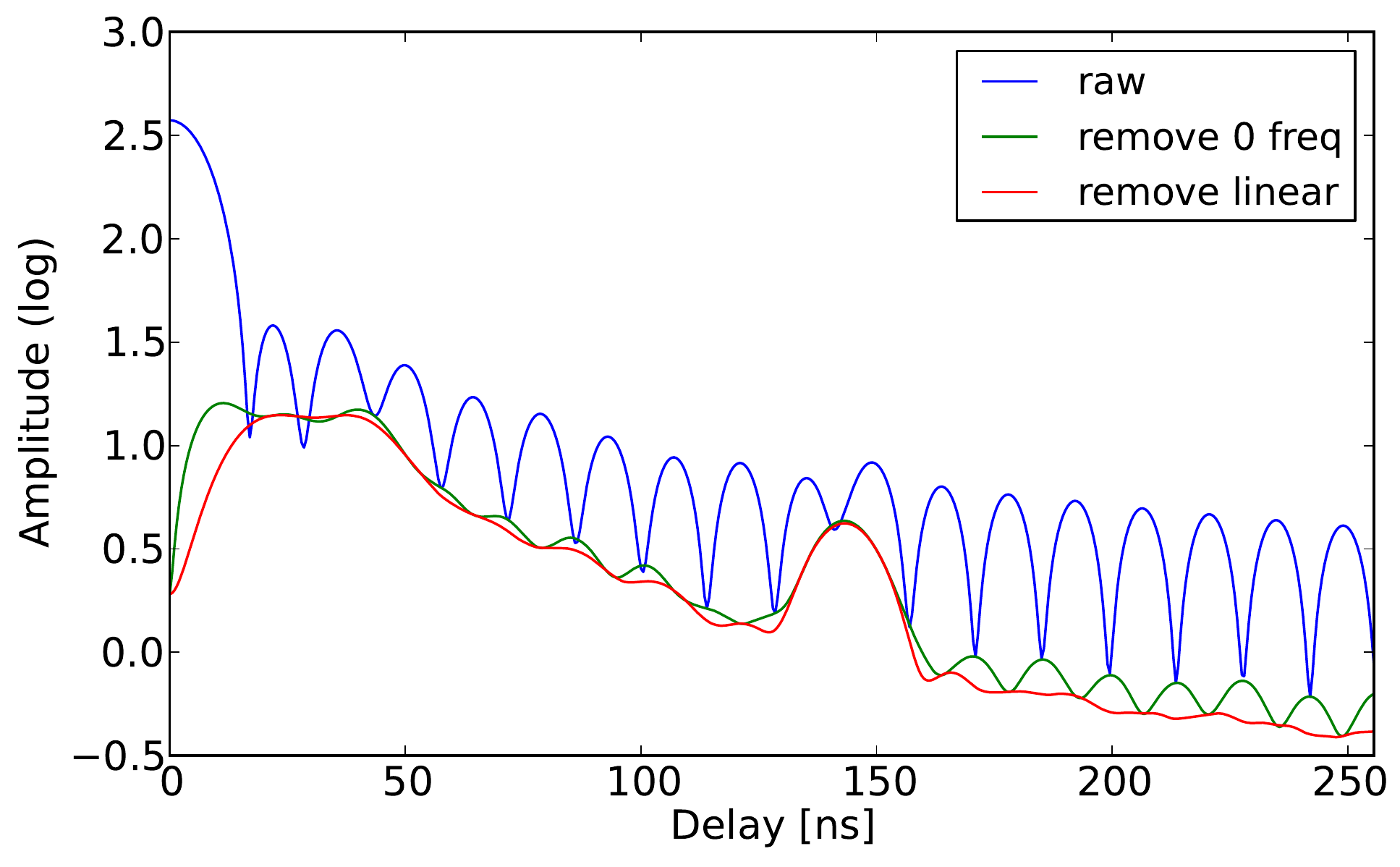}  \\
    \includegraphics[width=0.47\textwidth]{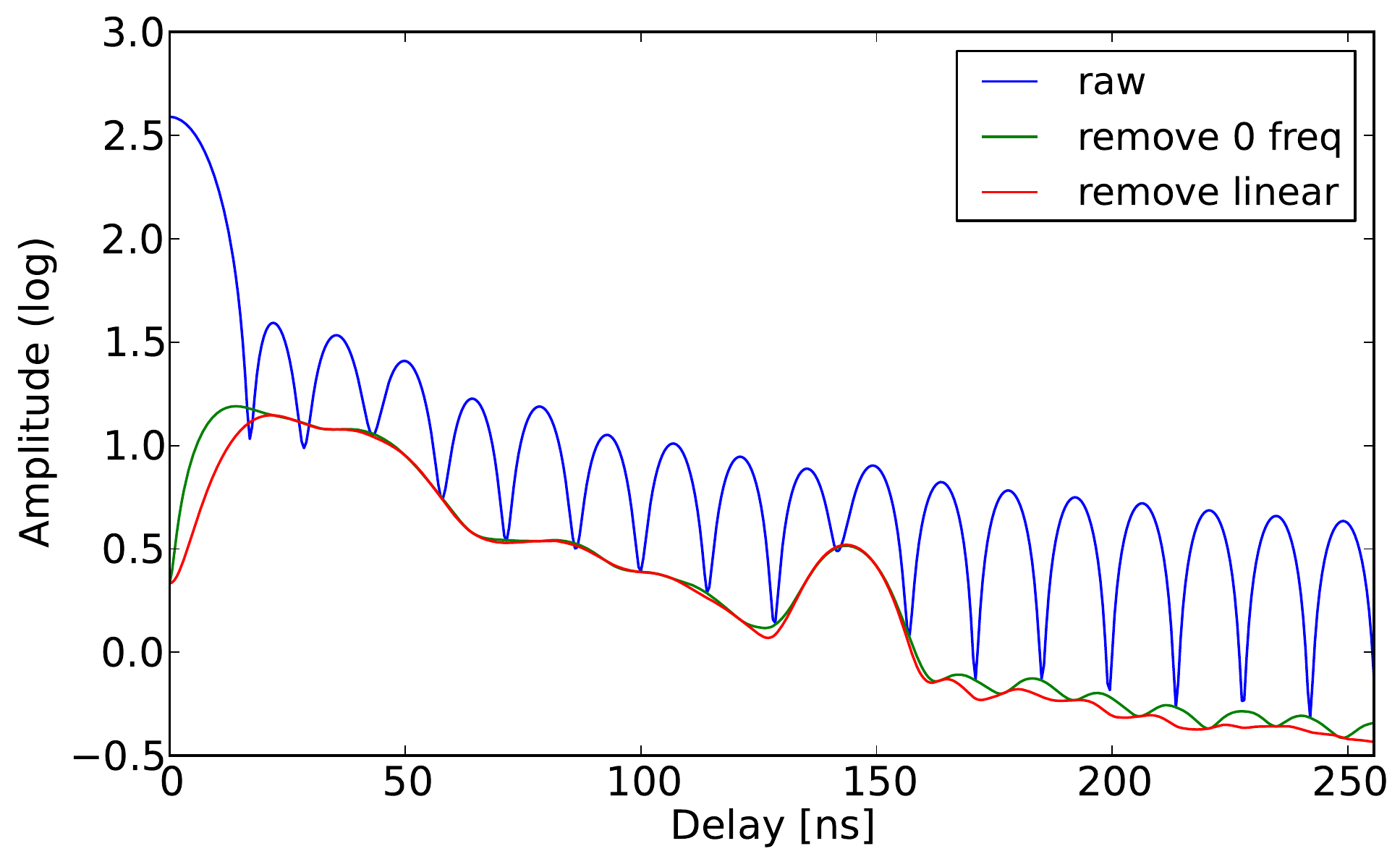}     
    \includegraphics[width=0.47\textwidth]{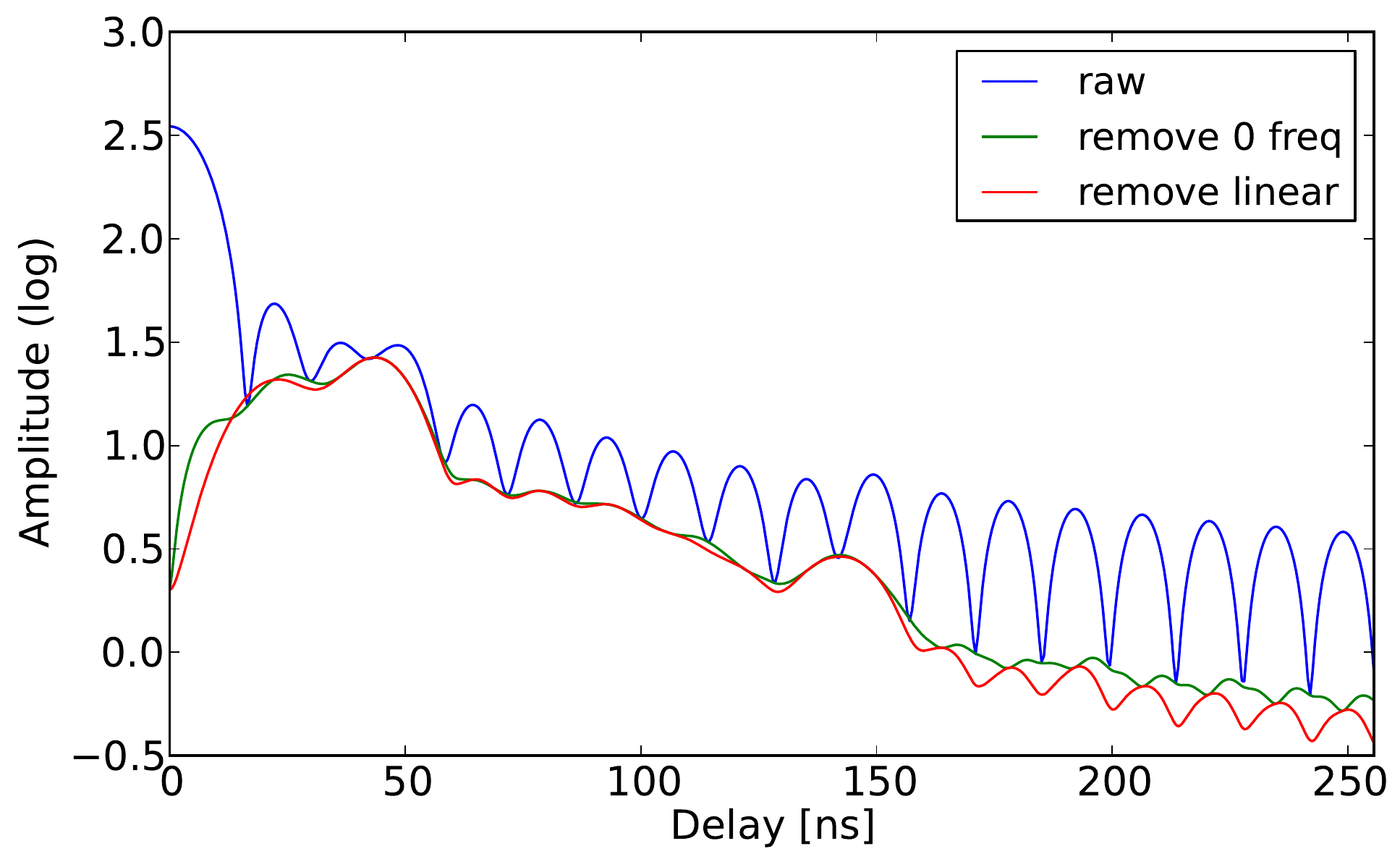}       
    \caption{Averaged time delays of all auto-correlations when different radio sources transit. Top left: night. Top right: Cyg A. Bottom left: Cas A. Bottom right: Sun.}
    \label{fig:mean_fft_cmpr_src}
\end{figure}

In Fig.~\ref{fig:mean_fft_cmpr_src}, we show the average delay spectra for these four cases. The raw spectra show strong oscillating side lobes, but these can be suppressed by removing the zero frequency component. In the spectra with the average and linear component removed, we see the low frequency peak and the slope to the right of it are somewhat different in each case, but the 142~ns peak exists in all cases. Interestingly, the peak at 142~ns becomes less prominent in the case of the Sun. The reflection coefficient of the feed cable should be largely independent of the signal strength, so if this peak appears to be lower, it actually means the low frequency peak is taller in this case because the frequency spectra are normalized. This is perhaps due to the stronger contribution from the reflection by the antenna for the case of the Sun.

\subsection{Verification Experiment}

In the above, we speculated that some of the features in the frequency spectra are associated with either standing waves on the antenna, or standing waves between interfaces in the signal chain. In particular, from the values of the delay, we associated the features with the feed cable (15~m) and IF cable (4~m in total, or 2~m if considered as two separate ones). We can test this speculation by experiments. 
A simple experiment is to change the cable length and check if the standing wave structure changes. 

\begin{figure}[htbp]
  \centering
  \includegraphics[width=0.48\textwidth]{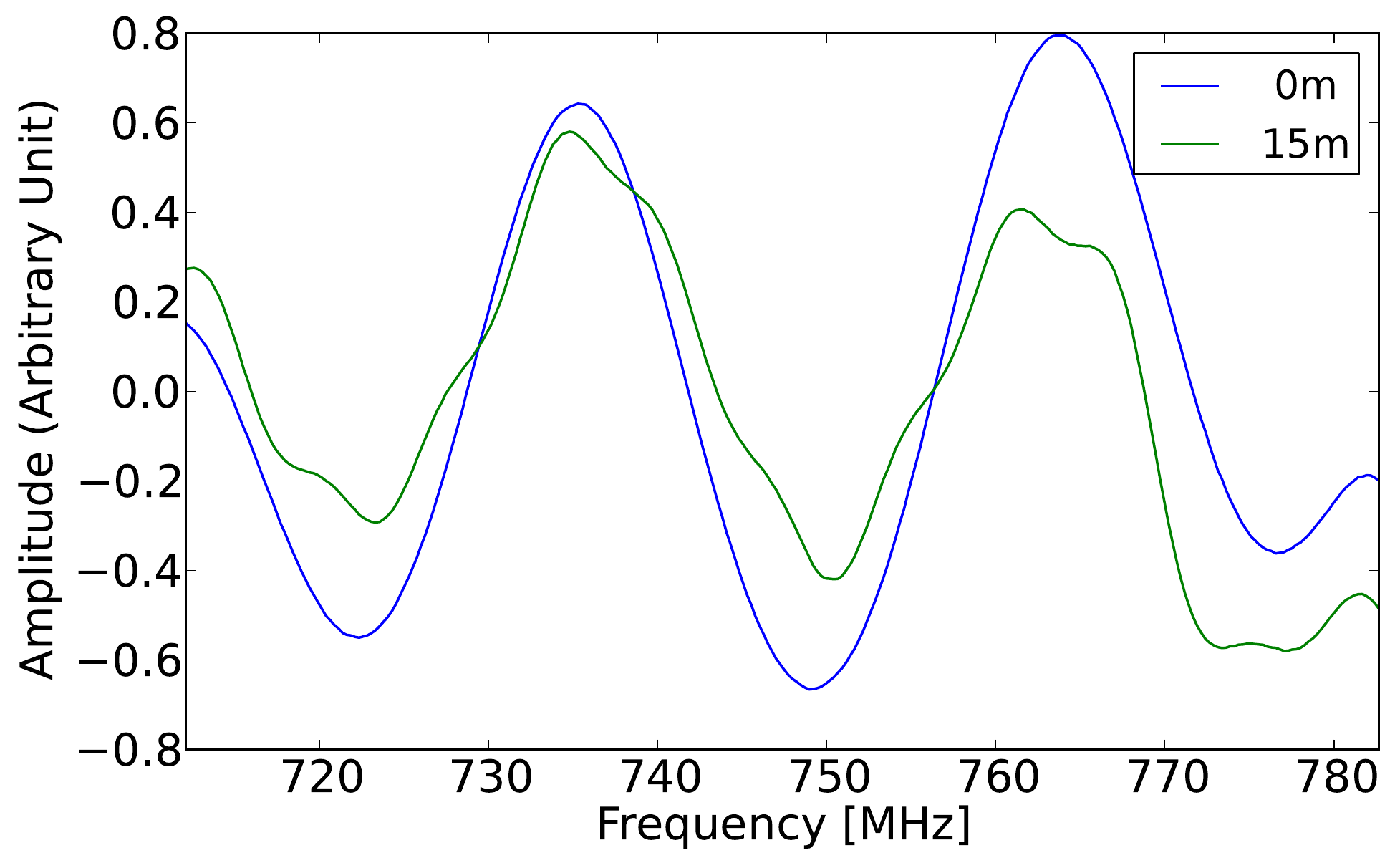}
  \includegraphics[width=0.48\textwidth]{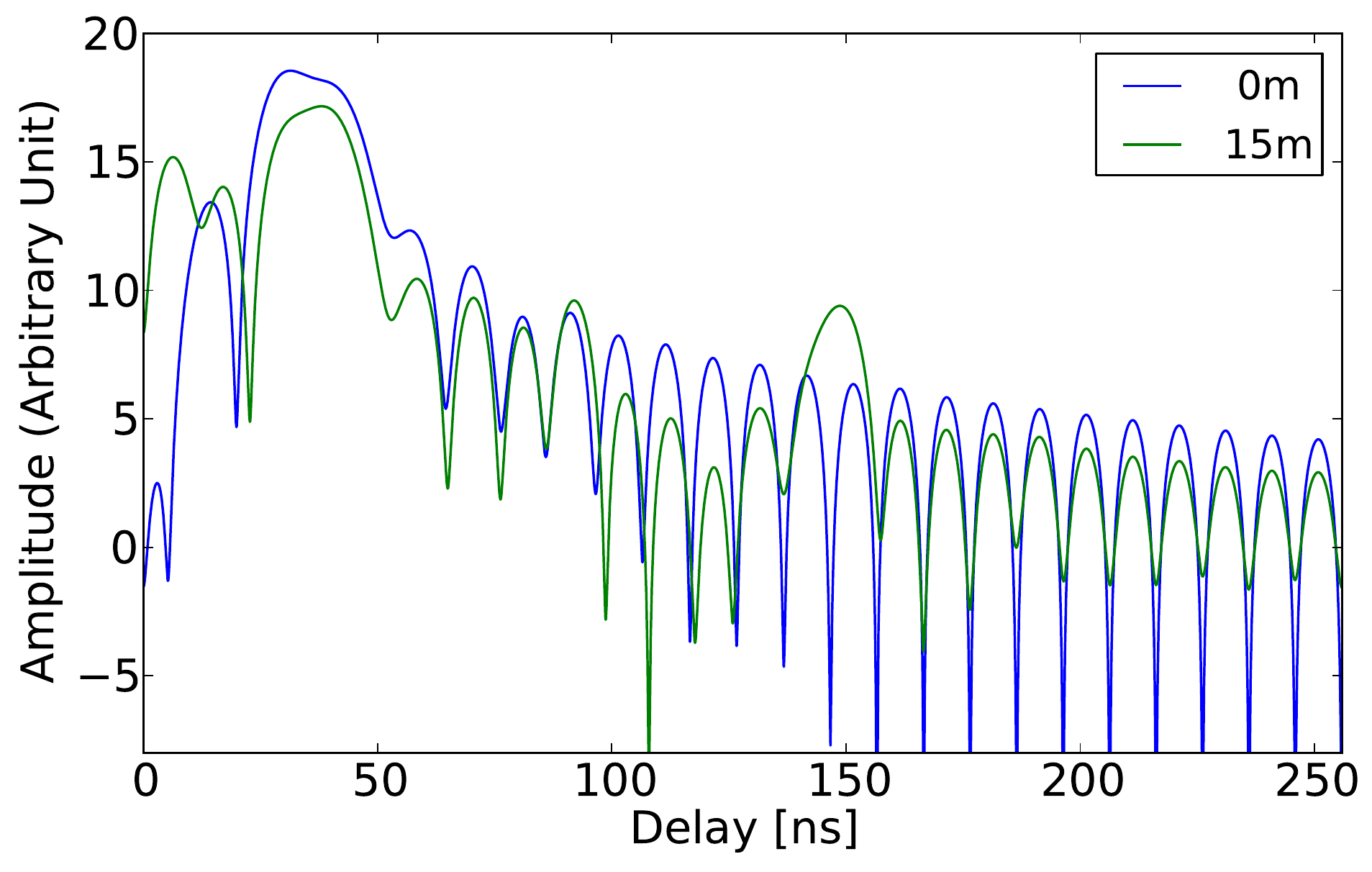}\\
  \includegraphics[width=0.48\textwidth]{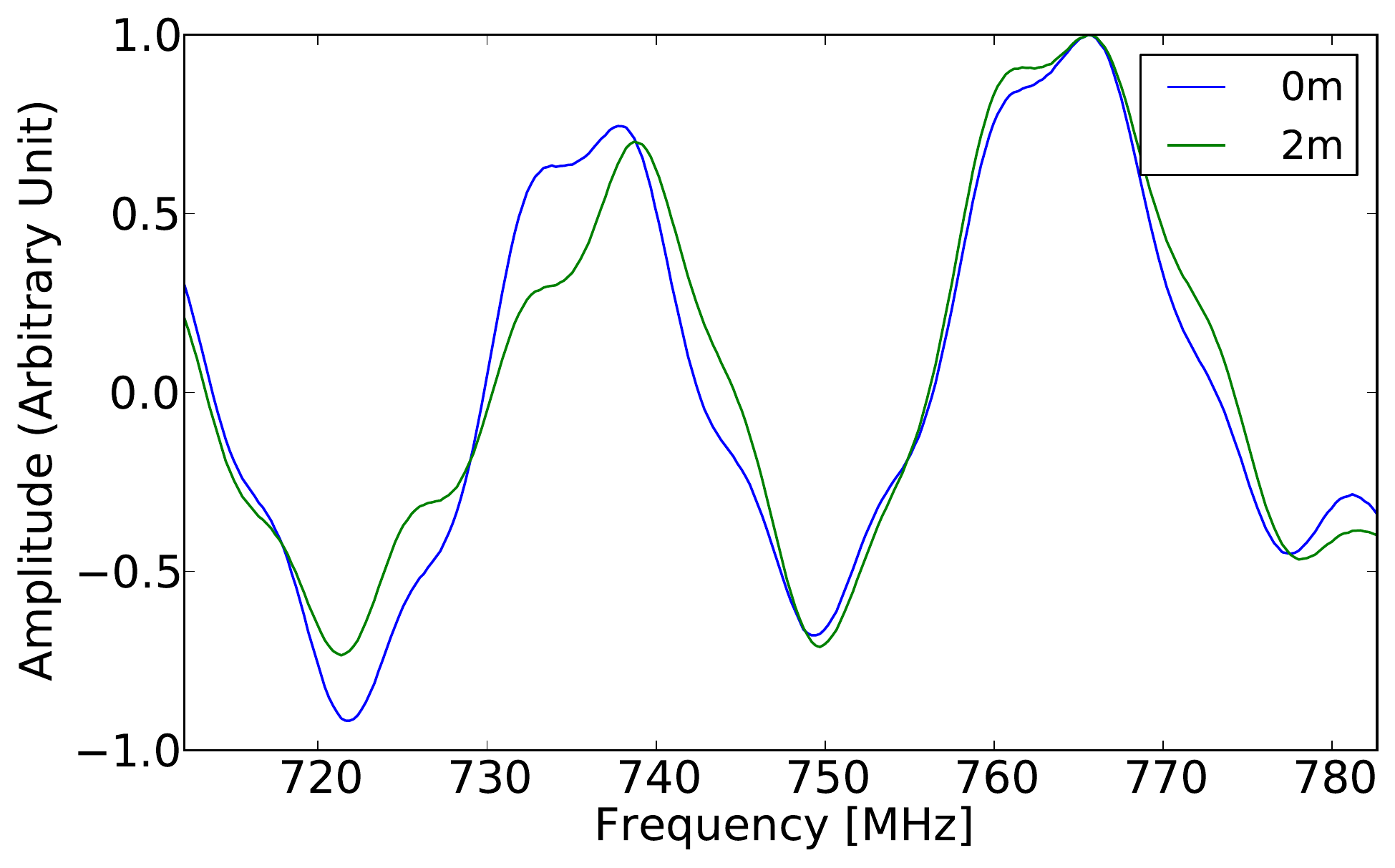}
  \includegraphics[width=0.48\textwidth]{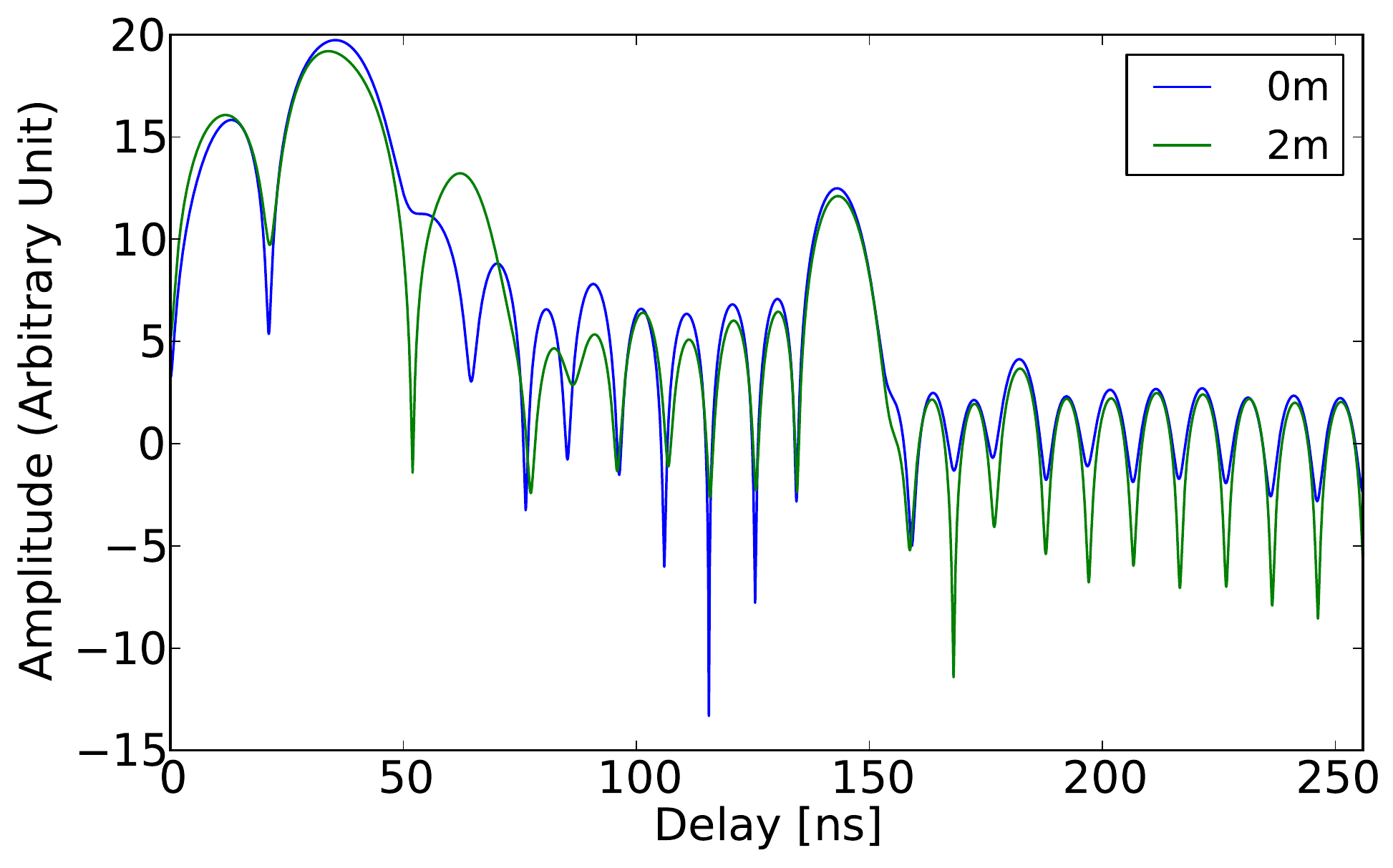}
  \caption{A comparison of the bandpasses and their delay spectra when cables of different lengths are inserted at different connection points. Top: the bandpass (left) and the delay spectrum (right) curves when a 15-meter feed cable is inserted behind LNA; ``0 m'' represents no insertion. Bottom: the bandpass (left) and delay spectrum (right) curves when a 2-meter cable is inserted behind the IF cable; ``0 m'' for no insertion. }
\label{fig:exp_cmpr}
\end{figure}

To check the standing wave effect in the feed cable, we did the following experiment: first, we replaced the feed antenna and its feed cable with a 50~$\Omega$ load resistor connected to the original LNA.  The output was connected directly to the optical transmitter and the rest of the system's signal chain. The auto-correlation spectrum is shown in the top left panel of Fig. \ref{fig:exp_cmpr} as the curve labeled ``0 m''. We then inserted a 15~m cable between the LNA and the optical transmitter;  the result is labeled ``15 m'' in the figure. The corresponding delay spectra are shown in the top right panel. The regular oscillations in this spectrum are the sidelobes of the Fourier transform window function, which we will ignore. What is relevant here is that when the load is directly connected into the signal chain, there is no strong peak at 142~ns except for the regularly spaced side lobes, but such a peak appeared in the delay spectrum when it is connected via the 15~m cable, thus proving that the 15-meter feed cable is indeed the origin of the $\sim$142~ns peak. 

To check the standing waves in the IF cables, we inserted a 2-meter cable between the existing IF cable and the correlator. The spectrum is shown in the bottom left panel of Fig. \ref{fig:exp_cmpr}, and the delay transform is shown in the bottom right panel. These curves are re-normalized to compensate for the attenuation of the inserted cable. A new peak at $\sim$ 60~ns appeared, consistent with a cable of 6~m total length, which would have a delay of 57~ns for a round trip with a wave speed of $0.7c$. Unfortunately, as the low frequency peaks blended together, it is not clear from this experiment whether there is a peak associated with the 2~m length cable and what its strength is. 

\begin{figure}[htbp]
  \centering
  \includegraphics[width=0.48\textwidth]{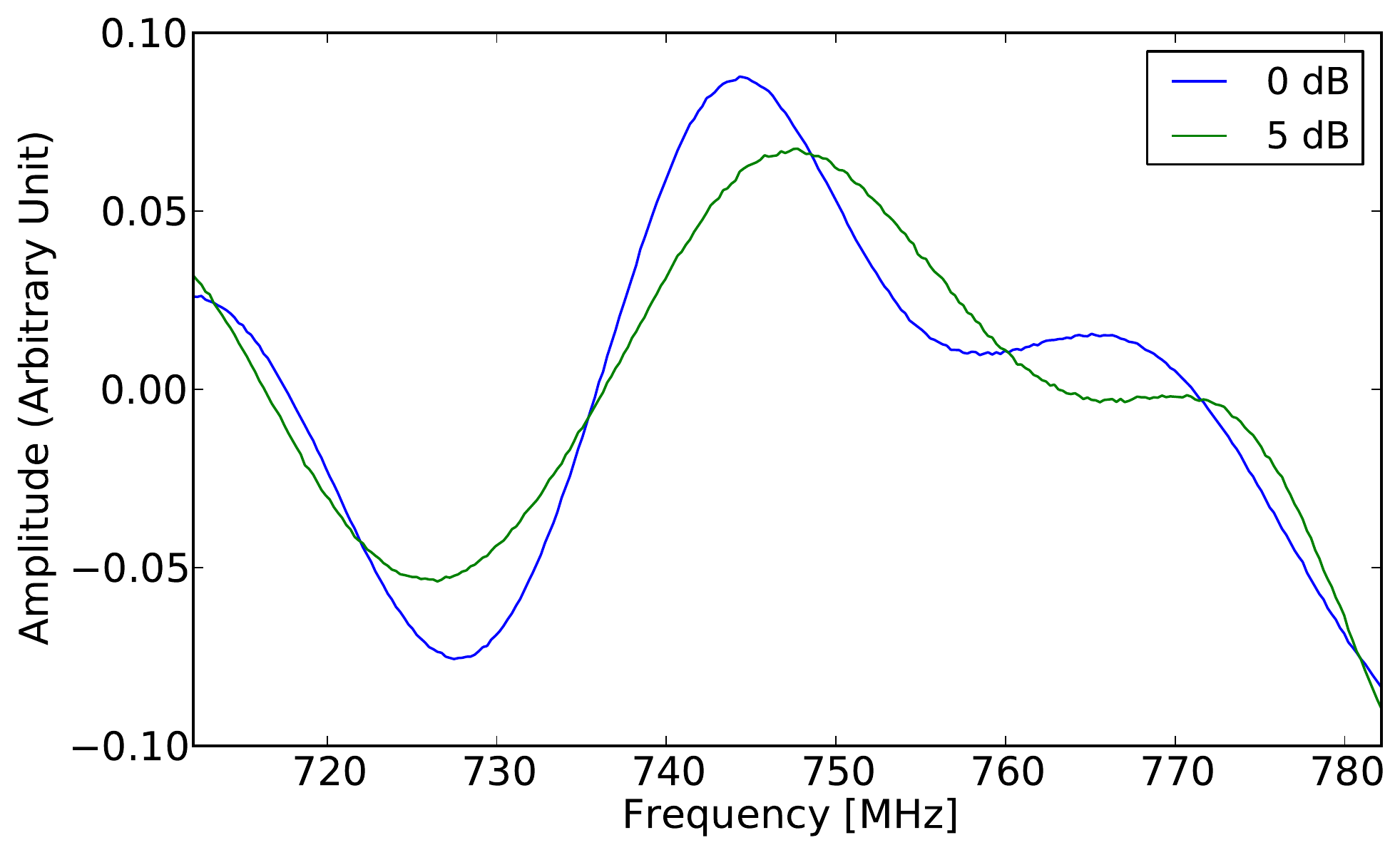}
  \includegraphics[width=0.48\textwidth]{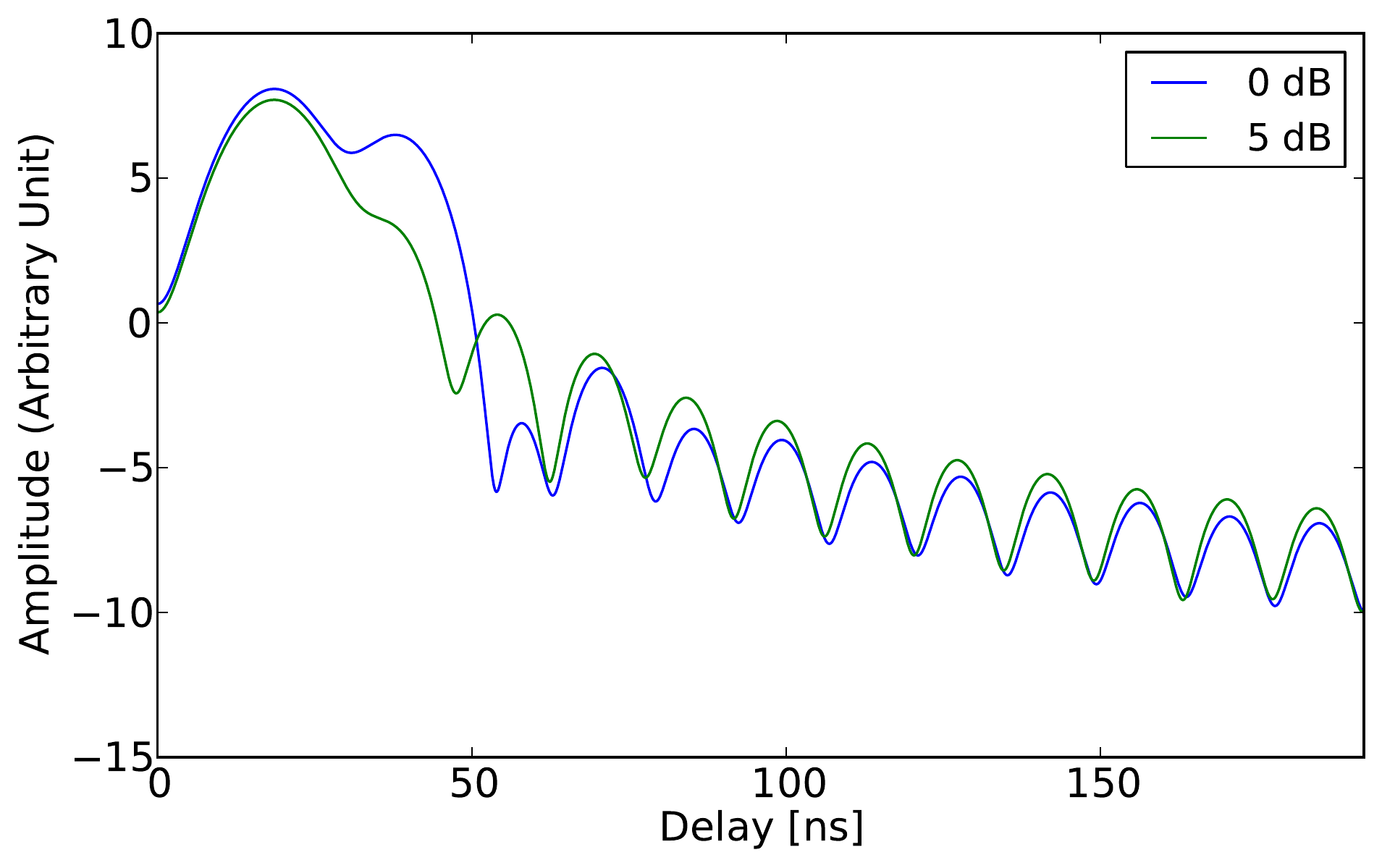}
  \caption{A comparison of the bandpasses (left) and their delay transforms (right) when a 5 dB attenuator is inserted in the IF cable part. ``0 dB'' for no insertion. The delay peak at $\sim$40~ns is partly mitigated after the 5 dB attenuator is inserted.}
  \label{fig:exp_atten_IF}
\end{figure}

Another experiment is to disconnect the antenna, using a 50 $\Omega$ load as input, and insert an attenuator into the signal chain. With the antenna disconnected, the standing wave on the antenna is eliminated. 
The reflected wave is attenuated twice while the incident wave is attenuated only once, so the standing wave in the circuit is also reduced. \footnote{There are also single direction microwave isolators which only suppress the wave in one direction, but at the low IF frequency, it is difficult to find ones with large enough bandwidth to cover our whole IF band, we therefore use an attenuator instead.}. 
We inserted a 5 dB attenuator between the existing IF cable and the correlator. The spectra are shown in the left panel of Fig. \ref{fig:exp_atten_IF}, and the delay transforms are shown in the right panel. The attenuation of the inserted attenuator is compensated by re-normalization.

We find that the amplitude of the wiggle has been mitigated a little. This is also shown clearly in the delay transform, where the peak at $\sim$ 40~ns is partly reduced. However, we note there are also variations, for some channels the change caused by the attenuator is less obvious, indicating less standing wave on the IF channel for these channels. We conclude that a part of the peak at low delay is due to the standing wave in the IF cable part. However, a broad peak at low delay (below 50 ns) remains, showing that there maybe other source or origin for this broad peak.

\subsection{Correction of Reflection Effects}
\label{sec:correction}

If the wiggles in the response are produced by reflections and standing waves, it is possible to make a correction once it is measured. We follow a reflection parameter estimation procedure similar to \citet{Kern:2019ytc}. First, make a delay transform of the frequency spectrum of the auto-correlation visibility. In doing so, we zero-pad the auto-correlation in frequency space and apply a window function to minimize the side-lobe power. With the delay spectrum, we make an initial estimate of the reflection delay $\tau$ and amplitude $A$ as $\tau_{0} = \tau_{\text{peak}}$, and 
$A_{0} = |\tilde{V}(\tau= \tau_{\text{peak}})| / |\tilde{V}(\tau = 0)|$. 
We eliminate the peak of reflection by $V^{\text{cal}} = V (1+\epsilon)(1+\epsilon^*)$,
where $\epsilon = A e^{i[2\pi \nu \tau + \phi]}$. We also refine the parameters $A, \tau, \phi$ to minimize the residual peak. If there are several peaks due to reflections in $|\tilde{V}|$, these can be eliminated by iteration, with the highest peak eliminated first, then the second peak, and so on. One can also make a joint fit of all parameters.

\begin{figure}[htbp]
  \centering
  \includegraphics[width=0.47\textwidth]{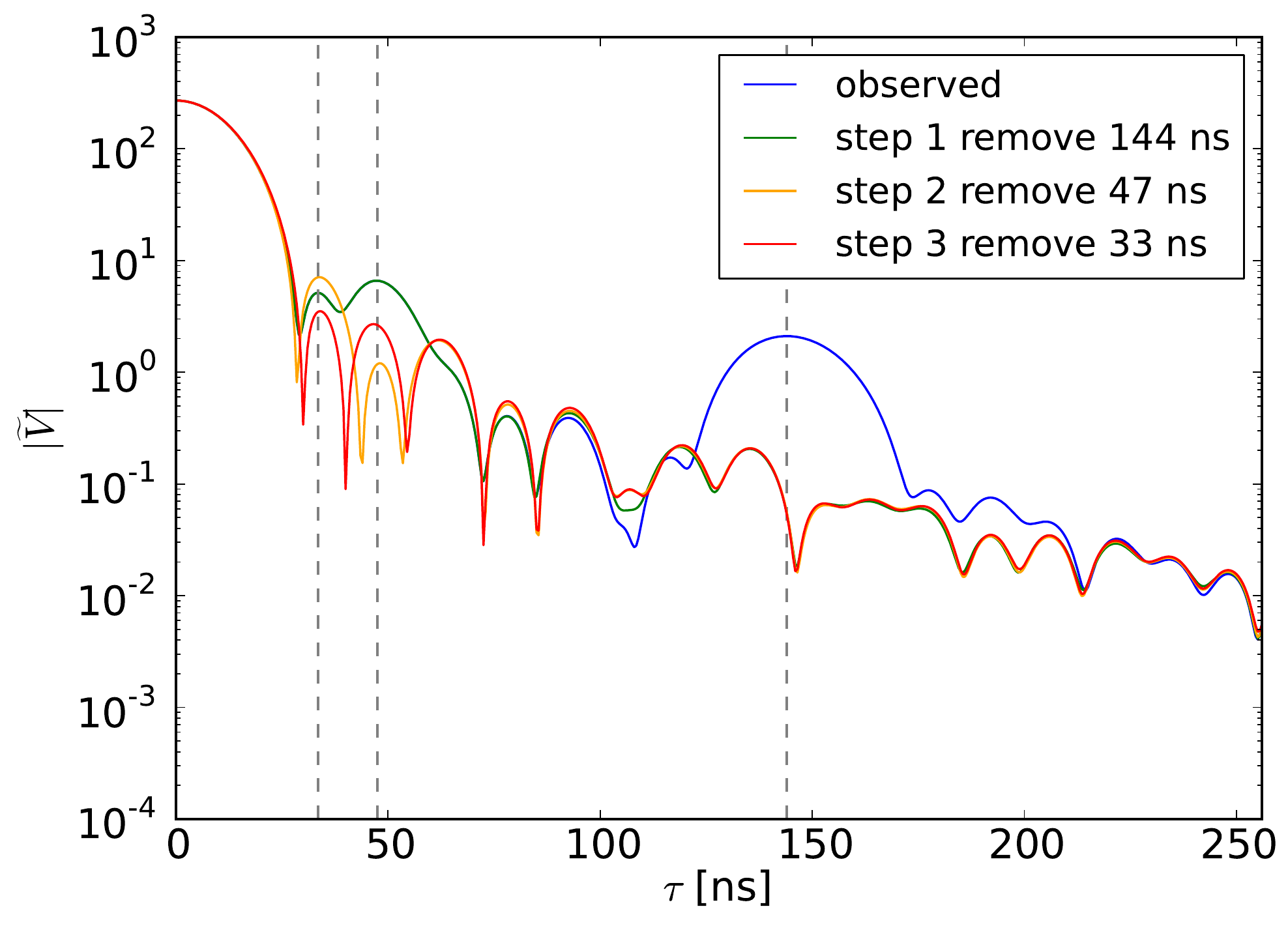}
  \includegraphics[width=0.47\textwidth]{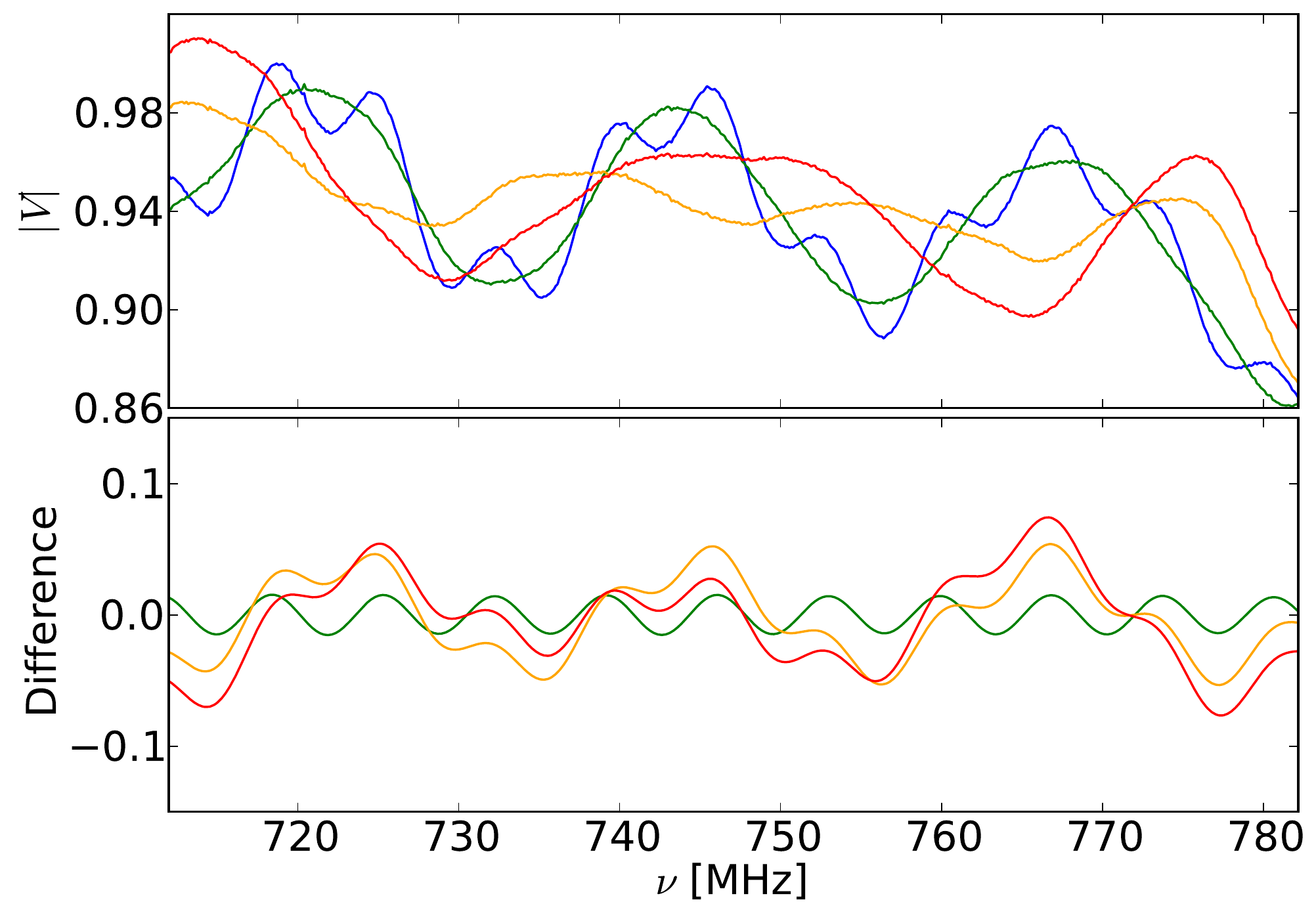}
  \caption{The reflection correction procedure applied to the signal channel A2Y. Left: the delay spectra of the observed visibility before and after correction. Top right: the frequency spectra before and after correction. Bottom right: the correction. Three peaks are removed from the original delay spectrum indicated by vertical dashed lines. The corresponding correction results are shown with different colored curves, each labeled as ``step 1'', ``step 2'' and ``step 3''. }
  \label{fig:reflection}
\end{figure}

As an example, Fig.~\ref{fig:reflection} shows the correction being applied to the signal channel A2Y, where the corrected $V$s after each step are marked. The left panel shows the delay spectra before and after the correction. The top right panel shows the frequency spectra, while the bottom right panel shows the correction. In this case three peaks are identified in this spectrum, with the fitted parameters 
$A_{1} = 7.80 \times 10^{-3}, \tau_{1} = 144.6~\mathrm{ns}, \phi_{1} = 7.19~\rad$;
$A_{2} = 2.12 \times 10^{-2}, \tau_{2} =  45.6~\mathrm{ns}, \phi_{2} = 0.31~\rad$; and 
$A_{3} = 1.37 \times 10^{-2}, \tau_{3} =  31.0~\mathrm{ns}, \phi_{3} = 8.30~\rad$, respectively. 
Here, we firstly remove the clear peak at $\sim$144 ns. However, the latter two peaks partly overlap with each other.
The measured $A$ values seem to be reasonable for reflections at the various interfaces. The maximum strength of the reflected signal is at the level of $10^{-2}$, so there is not much difference whether one uses the single reflection fit based on Eq. \ref{eq:one-reflection} or the multiple reflections fit based on Eq. \ref{eq:multi-reflection}. 

In this figure, the corrections are applied one at a time, so that the correction effect can be seen clearly for each standing wave component. We can see that after step 1, which removed the peak at 144~ns, the rapid oscillatory modulation is suppressed, and we obtained a smoother frequency spectrum. After step 2, which removed the peak at $\sim$45~ns, the modulation is further reduced. However, it seems that step 3 which removes the peak at $\sim31$~ns does not make much improvement, perhaps because this peak overlaps with the 45 ns peak. Also, there remains some modulations on the whole observed frequency range even after removing these peaks.

\section{Conclusions}
\label{sect:conclusion}
In the analysis of the Tianlai cylinder array data, we found regular wiggles in the frequency spectrum. The presence of such wiggles can complicate the task of foreground subtraction in 21~cm observations. In this paper, we investigated the origin of such wiggles in the bandpass by using the delay transform, and we found that at least some are the result of reflections and standing waves in the antenna and signal chain of the telescope. We focus on the auto-correlation visibilities in this paper, as the auto-correlations are  simpler than the cross-correlations, quite suitable for a first approach to the problem.     

After analysis of the data, we found that although the bandpass appears to be different for each channel, the delay spectra show that they could have a common origin: standing waves at some interfaces in the instrument. The most clearly identified is the one with a time delay at $\sim 142 \ns$, which is generated by the 15-meter feed cable. This is also confirmed by direct experiment. We also found that there are modulations with a time delay around $\sim 40 \ns$, and it is thought to be a mixture of a standing wave on the antenna between the feed and the reflector, which has a time delay of $\sim 32 \ns$, and a standing wave in the IF cable, which may have 19~ns delay (for the 2 meter segment) and 38~ns delay (for the total 4 meter length).

We also found that the standing waves due to the background noise can be different from those induced by radio sources. Most noise is generated within the electronic circuits, while for radio sources the standing waves between the feed and reflector can be significant. This is most clearly seen in the case of the Sun, which has a signal strength comparable with the receiver noise in the  auto-correlation visibilities. Indeed, in careful analysis, we find that the standing wave pattern is different for sources from different directions, e.g. the Sun, Cyg~A, and Cas~A each produces different standing waves. Such a direction-dependent effect could pose a serious challenge to the high precision calibration of the array.

Once the standing waves are recognized, we may make corrections by removing the corresponding components. We find that at least for the clearly identified component at $\sim142$~ns, the oscillating modulations in the frequency spectrum could be mitigated, resulting in a smoother spectrum. For the observed modulations of lower delay values, the same procedure can be applied, but it is less clear how good the correction is because there are probably several standing waves mixed together here. Fortunately, 
the low delay standing waves generate modulations which vary more slowly with frequency, and thus have less impact on the 21~cm signal extraction. 

The present study is still limited in its scope. We have considered here only the auto-correlations, and treated each individual feed and circuit separately. In fact, neighboring feeds are coupled with each other, and such coupling may generate collective modes in standing waves. Indeed, we find some evidence of this even in the present analysis (e.g. for the feather-like feature in the delay spectrum induced by the Sun and Cas~A). Also, not all of the oscillatory features are fully explained. For example, our current analysis shows relatively small reflection coefficients, but in some visibilities, we can see quite strong modulations. These more complicated system effects due to coupling between the feeds will be studied in future work.   

\begin{acknowledgements}
The Tianlai cylinder array is operated with the support of NAOC Astronomical Technology Center. The Tianlai cylinder is built with the support of the Ministry of Science and Technology (MOST) grant 2012AA121701, and its survey is supported by MOST grant 2016YFE0100300, the National Natural Science Foundation of China (NSFC) grants 11633004 and 11473044, and the Chinese Academy of Sciences (CAS) grants QYZDJ-SSW-SLH017. 
The data analysis work is partially supported by the MOST grant 2018YFE0120800, National Key R\&D Program 2017YFA0402603, and the CAS Interdisciplinary Innovation Team (JCTD-2019-05). Part of the computations are performed on the Tianhe-2 supercomputer (with the support of NSFC grant U1501501) and the Tianhe-1 supercomputer. 
Work at UW-Madison is partially supported by NSF Award AST-1616554. 
Authors affiliated with French institutions acknowledge partial support from CNRS (IN2P3 \& INSU), Observatoire de Paris and from Irfu/CEA. 
\end{acknowledgements}

\appendix                  

\bibliographystyle{raa}
\bibliography{refs}

\label{lastpage}

\end{document}